\documentclass[hyper(*)]{JHEP3}

\usepackage{epsfig,multicol,bbm}
\usepackage{enumerate}
\usepackage{bibentry}
\usepackage{amsmath}
\title{ M5 brane and four dimensional $\mathcal{N}=1$ theories I}

\author{Dan Xie

\\ School of Natural Sciences, Institute for Advanced Study \\
Princeton, NJ 08540, USA}

\abstract{Four dimensional $\mathcal{N}=1$ theories are engineered by compactifying six dimensional $(2,0)$ theory on a Riemann surface with regular punctures. A generalized Hitchin's
equation involving two Higgs fields is proposed as the BPS equation for $\mathcal{N}=1$ compactification. 
The puncture is interpreted as the singular boundary condition of this equation, and 
regular puncture is shown to be labeled by a nilpotent commuting pair. In this paper, we focus on a subset of regular puncture which is 
 described by rotating branes representing $\mathcal{N}=2$ puncture.
As an application, we show that
the Seiberg duality of $SU(N)$ SQCD with $N_f=2N$ and certain superpotential term is realized as different degeneration limits of the same punctured Riemann surface, and also find four more  dual theories.}

\begin{document}

\section{Introduction}
Gauge theory can be engineered using brane systems of type II string theory from which various physical properties can be understood nicely from brane configurations \cite{Giveon:1998sr},
 moreover, the lift of the above
brane configuration to M theory can usually lead to remarkable results about the IR behavior of the gauge theory. The above strategy has been successfully 
implemented  for  4d  $\mathcal{N}=2$ \cite{Witten:1997sc} and $\mathcal{N}=1$ theory \cite{Witten:1997ep,Hori:1997ab}, and in both cases all the relevant type II branes become a single M5 brane
and the IR behavior is controlled by a Riemann surface on which M5 brane wraps. 

The type II brane construction usually involves D brane ending on NS brane, and two dimensional conformal field theory description is singular at
the intersection, therefore, many questions about the UV theory, say S duality of 4d $\mathcal{N}=2$ theory, is not easily understood using type II branes since typically one 
need to move the branes to pass each other, and there are various difficult phase transition questions in this process.

Instead, one could try to engineer UV theory by directly starting with $N$ multiple M5 branes, and compactify it on a Riemann surface with co-dimensional
two defects (punctures), which basically represent various intersected branes. 
Gauge coupling is understood as the complex structure of the Riemann surface (as the gauge coupling is interpreted as the relative positions of the intersected branes \cite{Witten:1997sc}), 
and now nothing is singular in moving around the punctures to change the gauge coupling, 
so S duality of $\mathcal{N}=2$ theory \cite{Argyres:2007cn} is manifest in this representation \cite{Gaiotto:2009we}. 

More generally, one could engineer new four dimensional $\mathcal{N}=2$ theory as follows: 
simply compactify 6d $(2,0)$ theory on  a Riemann surface with various type of defects \footnote{The history of constructing four dimensional field theories 
using M5 branes in the case of no defects goes back to \cite{Maldacena:2000mw,Maldacena:2000yy}, see also \cite{Gauntlett:2003di} for constructing lower dimensional field theories using M5 branes.} .
Usually, the most important ingredients are the local property of the defects, which basically provide all the richness of M5 brane engineering. 
The important tool is the Hitchin's equation defined on the Riemann surface: various defects 
are singular boundary conditions to Hitchin's equation.
Using regular defects \footnote{$A_{N-1}$ regular defects 
are classified in \cite{Gaiotto:2009we}, $D_N$ regular defects are classified in \cite{Tachikawa:2009rb}, other type of regular defects including twisted lines are discussed in \cite{Tachikawa:2010vg,Chacaltana:2012zy,Chacaltana:2012ch}.},
 one can find lots of generalized superconformal 
quiver gauge theory \cite{Gaiotto:2009hg,Nanopoulos:2009uw,Nanopoulos:2010ga,Chacaltana:2010ks}. Using the irregular singularity which is classified in \cite{Xie:2012hs} for $A_{N-1}$ case, one can engineer many new Argyes-Douglas and 
asymptotical free theories. 

Such M5 brane engineering is also extended to four dimensional $\mathcal{N}=1$ theories \cite{Maruyoshi:2013hja,Benini:2009mz,Bah:2011je,Bah:2011vv,Bah:2012dg,Beem:2012yn,Gadde:2013fma,Maruyoshi:2013hja}, in particular, the global aspects of 
the compactification is found in \cite{Bah:2011vv}: Two line bundles $L_1$ and $L_2$ such that $L_1\bigotimes L_2=K$ ( $K$ is the canonical bundle) are defined on Riemann surface.
However, two important ingredients are still missing: first, the analog of Hitchin's equation is not found in the literature, and secondly various 
defects which provide all the richness are not discussed in a systematical way. Because of the above two missing pieces, the matter system for $\mathcal{N}=1$ theory is basically $\mathcal{N}=2$
matter system. 

Our main purpose in this paper is to fill in the above two gaps.
We propose a generalized Hitchin's equation for $\mathcal{N}=1$ compactification: 
\begin{align}
&D_{\bar{z}} \Phi_1 =D_{\bar{z}} \Phi_2=0, \nonumber\\
&[\Phi_1, \Phi_2]=0, \nonumber\\
&F_{z\bar{z}}+[\Phi_1, \Phi_1^*]ss^{*}+[\Phi_2, \Phi_2^{*}]tt^{*}=0,
\end{align}
here $\Phi_1$ and $\Phi_2$ are the sections of the line bundles $L_1$ and $L_2$ respectively, and they also transform in adjoint representation of the gauge group.
Notice that because $\Phi_1, \Phi_2$ are not the sections of cotangent bundle, we use fixed section $s\in K\bigotimes L_1^{-1}$ and $t\in K \bigotimes L_2^{-1}$ to make 
the last equation coordinate invariant ($s^*$ and $t^*$ are sections of the dual bundle which
can be fixed using the hermitian metric of $L_1$ and $L_2$.) 
This equation is conformal invariant and therefore one could find solution on compact Riemann surface. 

The local regular singular solutions can then be easily found (for the local solution, we can ignore the fixed section):
\begin{align}
&\Phi_1={e_1\over z},~~\Phi_2={e_2\over z} \nonumber,~~A_{\bar{z}}={h_1\over \bar{z}}+{h_2\over \bar{z}}, \nonumber\\
&[e_1, e_2]=0,~~~[h_1,h_2]=0, \nonumber\\
&[h_1,e_1]=e_1,~~[h_2,e_2]=e_2,
\end{align}
here  $e_1, e_2$ is taken to be nilpotent. Therefore the regular singularity is specified by the orbit of a nilpotent commuting pair, which is 
studied by Ginzburg in \cite{ginzburg2000principal}. When one of the nilpotent element is zero, we get the usual $\mathcal{N}=2$ regular puncture.

\begin{center}
\begin{figure}[htbp]
\small
\centering
\includegraphics[width=10cm]{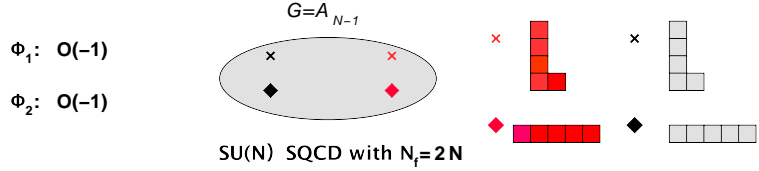}
\caption{The  $M5$ brane compactification data for $SU(N)$ SQCD with $N_f=2N$.  Black Young Tableaux means that $\Phi_1$ is singular 
at the puncture while $\Phi_2$ is zero, and the similar interpretation applies to the red Young Tableaux.}
\label{intro}
\end{figure}
\end{center}

In summary, Our theories are derived by starting with six dimensional $(2,0)$ theory and compactify it on a Riemann surface with the following data:
\begin{itemize}
\item A punctured Riemann surface $M_{g,n}$.
\item A rank two line bundle $L_1\bigoplus L_2$ such that $\text{deg}(L_1)+\text{deg}(L_2)=\text{deg}(K)$ with $K$ the canonical bundle, 
and two complex scalars $\Phi_1, \Phi_2$ are holomorphic sections of them.
\item The local puncture types: a commuting nilpotent pair.
\end{itemize}
We conjecture that in the IR the theory flows to a 4d $\mathcal{N}=1$ fixed point. See figure. \ref{intro} for the description of $SU(N)$ SQCD with $N_f=2N$.

\begin{center}
\begin{figure}[htbp]
\small
\centering
\includegraphics[width=12cm]{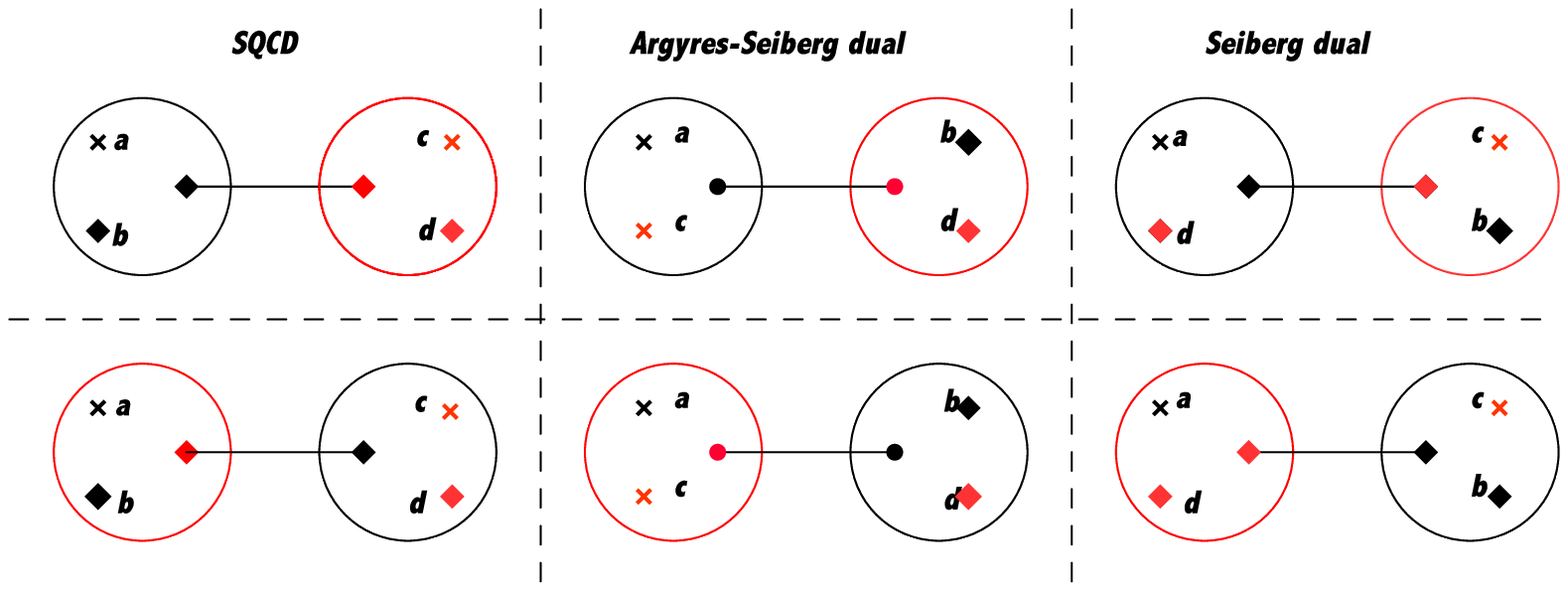}
\caption{Different duality frames of $SU(N)$ SQCD with $N_f=2N$, and the quartic superpotential couplings are the same
for the duality frames in the same column. }
\label{duality}
\end{figure}
\end{center}

In this paper, we focus on regular punctures which could be derived by rotating $\mathcal{N}=2$ puncture:  $\mathcal{N}=2$ puncture can be 
realized as the half-BPS boundary condition of $\mathcal{N}=4$ Super Yang-Mills (SYM) theory on a segment, and can be represented nicely by  D3-D5 system\footnote{Such descriptions are available for punctures of $A_N$ and $D_N$ theories \cite{Gaiotto:2008ak}.}. 
Similarly, $\mathcal{N}=1$ puncture can be realized as the quarter-BPS boundary condition of $\mathcal{N}=4$ SYM\footnote{Those quarter-BPS boundary
condition is studied in \cite{Hashimoto:2013}.}, and most simple
ones are formed by rotating some of  $\text{D}5$ branes to $\text{D}5^{'}$ branes \cite{Elitzur:1997fh,Elitzur:1997hc}, which could be labeled by a colored Young Tableaux.

As an application, we study $\mathcal{N}=1$ dualities using the Riemann surface picture, and 
we find that Seiberg duality \cite{Seiberg:1994pq} is also realized as different degeneration limit of the same Riemann surface much as the story of $\mathcal{N}=2$ S duality. We identify the complex structure moduli as the quartic 
superpotential term:
\begin{equation}
W=ctr(\mu_1\mu_2),
\end{equation}
here $\mu_1,\mu_2$ are the momentum map of two gluing punctures in the degeneration limit.  After doing Seiberg duality, we have $c\rightarrow -{1\over c}$, which matches the Riemann surface picture.
Such quartic coupling is exactly marginal in the case of $SU(N)$ with $2N$ flavor \cite{Leigh:1995ep} and we conjecture that it
is exactly marginal for the theory  considered above. This type of quartic superpotential is important for duality to work nicely in our picture: the matter in various duality frames can be represented by three punctured sphere.
Notice that for $\mathcal{N}=2$ gluing, there is a cubic superpotential term:
\begin{equation}
W=\tau tr \Phi(\mu_1-\mu_2),
\end{equation}
and again this coupling is identified as the complex structure moduli, and under S duality $\tau\rightarrow -{1\over \tau}$.
Therefore, $\mathcal{N}=1$ duality works in the same way as $\mathcal{N}=2$ S duality.
Various duality frames for $SU(N)$ SQCD are shown in figure. \ref{duality}. 
Similar dualities are discussed in \cite{Gadde:2013fma}, but in our description, matter system is represented by a three punctured sphere with black and red punctures and therefore admits
a M5 brane construction, it would 
be interesting to identify our matter system with theirs.

This paper is organized as follows: in section II, we discuss the basic ingredients of engineering $\mathcal{N}=1$
theory from M5 brane: a generalized Hitchin's equation and a classification of regular punctures. 
In section III, we discuss $\mathcal{N}=1$
duality for theories defined using 6d $A_{N-1}$ type theory and fully rotated punctures; in section IV, we study theories engineered using partial rotated puncture and $D_N$ theory. Finally, a conclusion is given in section V.

\section{ $\mathcal{N}=1$ compactification}
\subsection{Topological partial twist and global breaking to $\mathcal{N}=1$}
One can get lower dimensional supersymmetric field theory by doing partial topological twist of a higher dimensional field theory \cite{Bershadsky:1995vm}. 
In particular, one can do topological partial twist on 6d $(2,0)$ theory to get four dimensional $\mathcal{N}=1$ theory as discussed in \cite{Maldacena:2000mw,Anderson:2011cz}. Here let's give 
a brief review. The global symmetry groups of 6d $(2,0)$ theory are: the $Sp(4)$ R symmetry and $SO(6)\simeq SU(4)$ space-time symmetry. The supercharge $Q$ transforms under $SU(4)\times Sp(4)$ as
\begin{equation}
Q:~4\bigotimes 4,
\end{equation}
and the five adjoint scalars $\phi_i$ transforms  as
\begin{equation}
\phi:~1\bigotimes 5.
\end{equation}
 Let's consider a six manifold with product structure $R^4\times \Sigma$ so that the space-time symmetry group is decomposed as $SO(4)\times SO(2)$, and we use a R symmetry subgroup
 $U(1)_{45}\times U(1)_{89}\subset SO(5)$ to do the partial twist. 
 Before the twist, the supercharges transform under the above subgroup $SO(4)\times SO(2) \times U(1)_{45}\times U(1)_{89}$ as
 \begin{equation}
Q:~4\bigotimes 4\rightarrow ((2,1)_{1\over2}+(1,2)_{-{1\over 2}})\bigotimes (({1\over 2},{1\over 2})+({1\over 2},-{1\over 2})+(-{1\over 2},{1\over 2})+(-{1\over 2},-{1\over 2}))
 \end{equation}
 and the scalars decompose as
\begin{equation}
\phi:~1\bigotimes 5\rightarrow 1_0\bigotimes (({1\over2})_0+(-{1\over2})_0+0_{1\over2}+0_{-{1\over2}}+1_0).
\end{equation}
 
 Let's twist the theory by embedding the R symmetry into the $SO(2)$ symmetry 
 on Riemann surface: the new rotational symmetry group $SO(2)^{'}=SO(2)+a U(1)_{45}+b U(1)_{89}$. The  supercharges have charges under $SO(2)^{'}$: 
 \begin{equation}
 \pm{1\over2}\pm{a\over 2} \pm {b\over 2},
 \end{equation}
 so if $a+ b =1$, we have invariant supercharges. If $b=0, a=1$ or $b=1, a=0$,  one has eight unbroken supercharges, and we get four dimensional $\mathcal{N}=2$ theory whose
  $R$ symmetry group is $U(1)_{45}\times SU(2)$ and $U(1)_{89}\times SU(2)$, we call them $\text{NS}$ type theory and $\text{NS}^{'}$ type theory whose meaning 
 will be clear once we consider the type IIA brane configurations in next subsection. The scalars which have nonzero charges under $SO(2)^{'}$ are
 \begin{align}
&\Phi_1:~~~{1\over 2},~~~\Phi_2:~~~0,~~~~~~~~\text{NS}~\text{theory}, \nonumber\\
&\Phi_1:~~~~0,~~~\Phi_2:~~{1\over 2},~~~~~~~\text{NS}^{'}~\text{theory},
 \end{align}
here $\Phi_1$ and $\Phi_2$ is related to the original five scalars as $\Phi_1=\phi_4+i\phi_5$ and $\Phi_2=\phi_8+i\phi_9$. In our normalizations, $\phi_4$ and $\phi_5$ 
are vectors under $SO(2)^{'}$ (let's take NS theory as an example), so $\Phi_1$ is a section of canonical bundle of Riemann surface.
 
 If both $a$ and $b$ are nonzero and satisfy equation $a+ b =1$, there are four unbroken supercharges and we get 4d $\mathcal{N}=1$ theory. 
 The two complex scalars $\Phi_1$ and $\Phi_2$ have charges 
 \begin{equation}
 \Phi_1:~~{a\over 2},~~~~~\Phi_2:~~{b\over 2},
 \end{equation}
 so in general we have two charged complex scalars which are holomorphic sections of two line bundles:
\begin{align}
\Phi_1 \in \Omega^0(\Sigma, E\bigotimes L_1), ~~~\Phi_2\in \Omega^0(\Sigma, E\bigotimes L_2),
\end{align}
and $L_1\bigotimes L_2=K$, where $K$ is cotangent bundle of Riemann surface and $E$ is the adjoint representation of gauge group.

\subsection{Punctures and local breaking to $\mathcal{N}=1$}
As we saw in last subsection, the $\mathcal{N}=1$ compactification of $(2,0)$ theory introduces two line bundles whose tensor product 
is the canonical bundle of Riemann surface. In this subsection, We would like to discuss the classification of local regular punctures. In the 
case of $\mathcal{N}=2$ compactification, Hitchin's equation plays a crucial role in classifying the punctures; Following similar lines, we propose a generalized 
Hitchin's equation, and then use it to classify the regular puncture of $\mathcal{N}=1$ compactification.

\subsubsection{Hitchin's equation and $\mathcal{N}=2$ punctures}
Let's first review the regular punctures for $\mathcal{N}=2$ compactification.
The BPS equation for $\mathcal{N}=2$ compactification is the so-called Hitchin equation, which can be derived in
the following way: one further compactify four dimensional theory on $T^2$, and then consider the compactification in different order: first on $T^2$
and then on Riemann surface $\Sigma$. In the first $T^2$ compactification, one get four dimensional $\mathcal{N}=4$ SYM, and 
in the second step, we do topological twist on $\Sigma$(it turns out the twist is the GL twist studied in \cite{Kapustin:2006pk}) and find
the Hitchin's equation \cite{Kapustin:2006pk}. 

The local form of Hitchin's equation can be derived from dimensional reduction of four dimensional self-dual Yang-Mills equation:
\begin{equation}
F_{12}=F_{34},~~F_{13}=F_{42},~~~F_{14}=F_{23}.
\end{equation}
After dimensional reduction to two dimensions \cite{Hitchin:1987ab}, i.e. all the fields are independent of coordinates $x^3,x^4$, we have 
\begin{equation}
F_{12}=[\phi_3,\phi_4],~~D_1\phi_3=-D_2 \phi_4,~~D_1\phi_4=D_2\phi_3.
\end{equation}
Define $\Phi={1\over2}(\phi_3-i\phi_4)$ and use the complex coordinate $z=x^1+ix^2$, we have the familiar Hitchin's equation:
\begin{equation}
F_{z\bar{z}}+[\Phi, \Phi^*]=0,~~~D_{\bar{z}} \Phi= D_{z} \Phi^{*}=0.
\end{equation}
To make this equation coordinate invariant on a curved Riemann surface, it is suggested by Hitchin that the scalar field $\Phi$ should be a one form.
As we see from the topological twisting, there is indeed one scalar which lives in the cotangent bundle of the Riemann surface.

The regular puncture means the singular boundary condition of Hitchin's equation,  and various fields have the following form \cite{Gukov:2006jk}:
\begin{align}
&\Phi_z={e\over z}dz,~~~A_{\bar{z}}={h\over \bar{z}}, \nonumber\\
&[h,e]=e,
\end{align}
here $e$ is a nilpotent element whose orbit is labeled by a Young Tableaux $\text{Y}=[n_1, n_2, \ldots, n_s] $\footnote{Let's take $G=SU(N)$ for example; The regular punctures of D and E type group are similar, for more details, see \cite{Tachikawa:2009rb,Chacaltana:2012zy}.} with $\sum n_i=N$: $e$ can be put into the standard form using Jordan blocks with size $n_i\times n_i$. 
The moduli space of Hitchin's moduli space ${\cal M}_H$ is identified with the Coulomb branch 
of four dimensional theory compactified on a circle. ${\cal M}_{H}$ has a distinguished complex structure which is independent of the radius of compactification circle, and one can define
a spectral curve  \cite{Hitchin:1987bc} which is then identified with the Seiberg-Witten curve. The local Coulomb branch (the local contribution to ${\cal M}_H$) could be identified
as the nilpotent orbit specified by $e$.
 
The local Higgs branch is described by the moduli space of Nahm's equation on a semi-infinite segment $[0,\infty]$ whose boundary 
condition at $0$ is specified by a dual Young Tableaux $Y_D$ ($Y_D$ is derived by transposing Y, i.e. the rows of $Y_D$ are the columns of $Y$.):
\begin{align}
&Y=[n_1, n_2, \ldots, n_s], ~~~n_1\geq n_2\ldots\geq n_s~~~\text{Local Coulomb}\nonumber\\
&Y^D=[r_1, r_2, \ldots, r_t].~~~ r_1\geq r_2\ldots\geq r_t~~~~~\text{Local Higgs}
\end{align}
The Nahm's equation can be derived by further assuming  all the fields in Hitchin's equation to be independent of coordinate $x^2$, and the equation reads
\begin{align}
&D_1 \phi_2=[\phi_3,\phi_4],~~D_1 \phi_3=[\phi_4,\phi_2],~~D_1 \phi_4=[\phi_2,\phi_3].
\end{align}
Define $\alpha=(A_1-i\phi_2)$  and $\beta={1\over 2}(\phi_3+i\phi_4)$, and Nahm's equation becomes a complex and a real equation:
\begin{align}
&{d\beta\over ds}+[\alpha,\beta]=0,~~~~~~~~~~~~~~~~~~~~~~~\text{Complex} \nonumber\\
&{d(\alpha^*-\alpha)\over ds}+[\alpha, \alpha^*]+[\beta, \beta^*]=0,~~\text{Real},
\end{align}
here $s=x^1$. The complex equation is invariant under the complex gauge group $G_c$.
By standard argument, for one special complex structure of the moduli space, imposing real equation is equivalent as dividing the complex 
equation by $G_c$.

Let's study the Nahm's equation on semi-infinite line $[0,\infty]$, and it has the following singular solution 
\begin{equation}
\alpha={h_D\over s},~~~\beta={e_D\over s},
\end{equation}
with $e^D$ a nilpotent element labeled by Young Tableaux $Y^D$, i.e. the standard form has Jordan block $r_i\times r_i$;
 and $h$ is a semi-simple element which satisfies $[h^D,e^D]= e^D$.  Each such nilpotent orbit defines a $SU(2)$ homomorphism $\rho_{Y^D}: SU(2)\rightarrow SU(N)$; and
the commutant subgroup $H$ of $\rho$ is then identified as the flavor symmetry group
\begin{equation}
H=S[\prod_lU(p_h)),
\end{equation}
where $p_h$ means the number of columns with height $h$ in Young Tableaux $Y^D$. 

\begin{center}
\begin{figure}[htbp]
\small
\centering
\includegraphics[width=8cm]{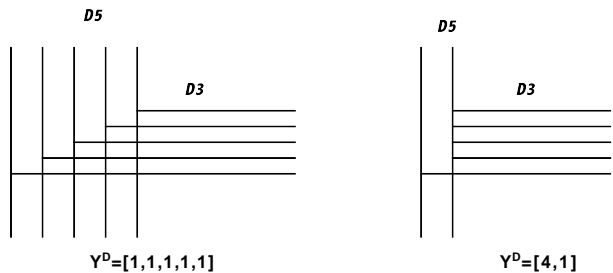}
\caption{The puncture is represented by half-BPS boundary condition of $\mathcal{N}=4$ SYM theory: left figure represents full puncture and right represents simple puncture. }
\label{nahm}
\end{figure}
\end{center}

This Nahm pole boundary condition specified by $Y^D$ can be nicely represented by a brane configuration of type IIB theory \cite{Gaiotto:2008sa}, here let's review it for the later use.  
The brane configurations are summarized in table. \ref{brane}. The above boundary condition is represented by a total of $t$ D5 branes, and the number of D3 branes suspended between
$i$th and $(i+1)$th D3 branes (counting from right to left) are 
\begin{equation}
N_i=N-(r_1+\ldots+r_i),
\end{equation}
see figure. \ref{nahm} for illustration. The boundary condition on the other side of the segment is trivial, and the local Higgs branch can be identified with the moduli space of 
Nahm's equation with trivial boundary condition at the infinity and Nahm's pole boundary condition specified by $Y^D$ at $s=0$.
\begin{table}
\begin{center}
  \begin{tabular}{ |l | c |c|c|c|c|c|c|c|c| r| }
    \hline
    ~&$x^0$&$x^1$ & $x^2$& $x^3$&$x^4$&$x^5$&$x^6$&$x^7$&$x^8$&$x^9$ \\ \hline
        D3&$\circ$&$\circ$ & $\circ$&~&~&~&$\circ$&~&~&~ \\ \hline
    NS5&$\circ$&$\circ$ & $\circ$& $\circ$&$\circ$&$\circ$&~&~&~&~ \\ \hline
        $\text{NS}5^{'}$&$\circ$&$\circ$ & $\circ$&$\circ$ &~&~&~&~&$\circ$&$\circ$ \\ \hline
    D5&$\circ$&$\circ$ & $\circ$&~&~&~&~&$\circ$&$\circ$&$\circ$ \\ \hline
    $\text{D}5^{'}$&$\circ$&$\circ$ & $\circ$&~ &$\circ$&$\circ$&~&$\circ$&~&~\\ \hline
  \end{tabular}
  \end{center}
  \caption{The brane configuration used in describing local $\mathcal{N}=2$ punctures.}
  \label{brane}
  \end{table}

\begin{center}
\begin{figure}[htbp]
\small
\centering
\includegraphics[width=12cm]{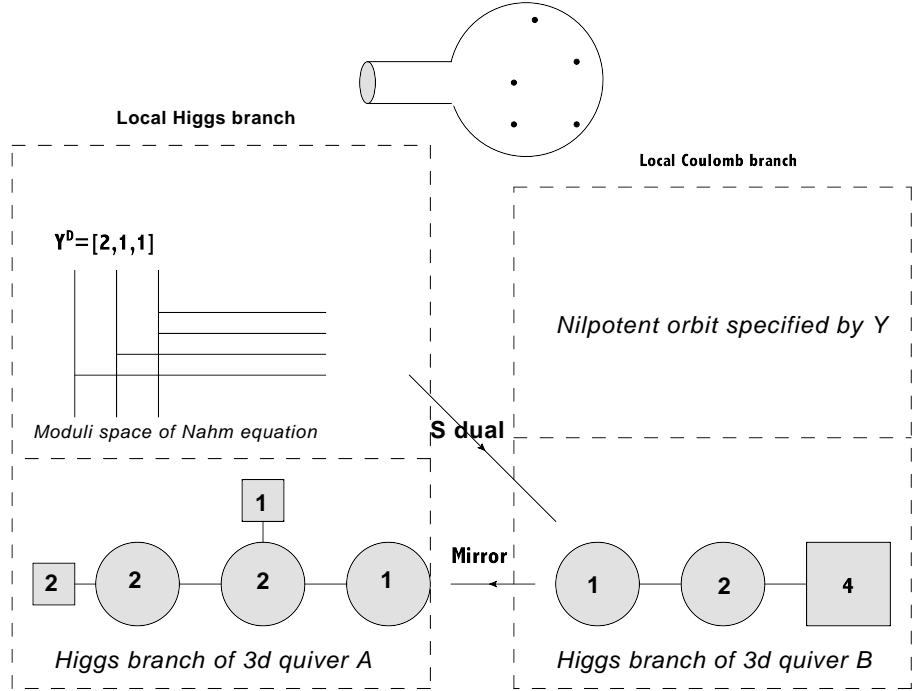}
\caption{Local Higgs and Coulomb branch of four dimensional $\mathcal{N}=2$ compactification can be identified with the Higgs branch of 3d quivers. }
\label{local}
\end{figure}
\end{center}

\begin{center}
\begin{figure}[htbp]
\small
\centering
\includegraphics[width=10cm]{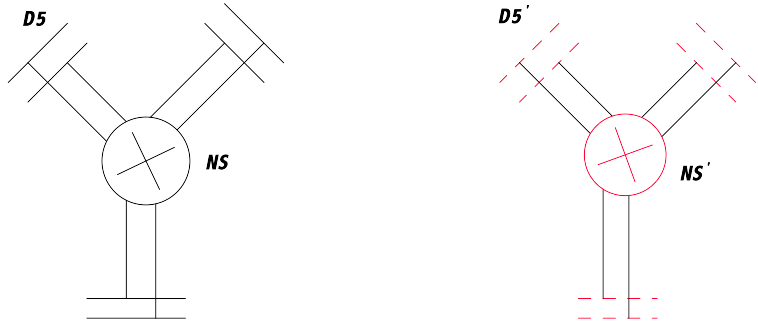}
\caption{NS and $\text{NS}^{'}$ type three sphere are represented by three junctions with specific boundary condition on three legs, and both lead to same $\mathcal{N}=2$ theory. }
\label{junction}
\end{figure}
\end{center}

By doing S-dual on above brane configuration, one find a three dimensional quiver $A$ whose Higgs branch gives the local Coulomb branch. The Higgs branch of three dimensional
mirror  $B$ of quiver $A$ describes the local Higgs branch, see figure. \ref{local}. Using the 3d quiver descriptions, we can see that the local Coulomb branch has a symmetry $SU(N)$ and 
local Higgs branch has a symmetry $H$.
The most basic ingredients for constructing 4d $\mathcal{N}=2$ theory using M5 brane
are the three punctured sphere which can be represented by a three junction  \cite{Benini:2010uu}. The Nahm's pole 
boundary condition is used to represent the puncture, see figure. \ref{junction}, which
is called a $\text{NS}$ type three sphere as we use $\text{D}5-\text{NS}$ branes. This type of three sphere preserves $U(1)_{45}$ symmetry.
Notice that we have the same story by replacing $\text{NS}5$ brane  and $\text{D}5$ brane with $\text{NS}5^{'}$ and $\text{D}5^{'}$ branes, which preserves  $U(1)_{89}$ symmetry, and it is called $\text{NS}^{'}$ type sphere. 
For the consideration of $\mathcal{N}=2$ theory, these two types of three spheres are completely equivalent as we only use one type of three spheres to preserve $\mathcal{N}=2$
supersymmetry. In the next subsection, when we consider $\mathcal{N}=1$ theory, the distinction between these two type of spheres becomes important.

\newpage
\subsubsection{Generalized Hitchin's equation and $\mathcal{N}=1$ punctures}
We would like to consider the BPS equations for $\mathcal{N}=1$ compactification. The derivation of the equation is similar to $\mathcal{N}=2$ case: one 
further compactify the theory on $T^2$ and then reverse the compactification order to first get four dimensional SYM theory; Using the Langrangian and 
topological twist of 4d SYM, one can get the desired equation. The local form of the equation is actually quite simple to get by using the $D4$ branes suspended between 
$D6$ and $D6^{'}$ branes as shown in Appendix A.   Here we derive the local equation as the dimensional reduction of higher dimensional equation: 
The $\mathcal{N}=1$ BPS equation can be derived from dimensional reduction of   6d self-dual equations \cite{Corrigan:1982th, Bak:2002aq}:
\begin{align}
&F_{13}=F_{24},~~~F_{14}=F_{32}, \nonumber \\
&F_{15}=F_{26},~~~F_{16}=F_{52}, \nonumber \\
&F_{35}=F_{46},~~~F_{36}=F_{54},\nonumber \\
& F_{12}=F_{34}+F_{56}.
\end{align}
Let's do dimensional reduction to two dimensions which has coordinates $x^1,x^2$, and we get:
\begin{align}
&D_1 \phi_3=D_2 \phi_4,~~D_1 \phi_4=-D_2\phi_3, \nonumber \\
&D_1 \phi_5=D_2 \phi_6,~~D_1 \phi_6=-D_2\phi_5, \nonumber \\
&[\phi_3, \phi_5]=[\phi_4,\phi_6]~~~[\phi_3,\phi_6]=[\phi_5,\phi_4],\nonumber \\
&F_{12}=[\phi_3, \phi_4]+[\phi_5,\phi_6]. 
\end{align}
Define $\Phi_{1}={1\over2}(\phi_3-i\phi_4)$ and $\Phi_{2}={1\over2}(\phi_5-i\phi_6)$, then the above equation becomes:
\begin{align}
&D_{\bar{z}} \Phi_1=0,~~~~D_{\bar{z}} \Phi_2= 0, \\
&[\Phi_1, \Phi_2]=0,\\
& F_{z\bar{z}}+[\Phi_{1}, \Phi_1^*]+[\Phi_2,\Phi_2^*]=0.
\end{align}
As we learn from 4d $\mathcal{N}=1$ twist, there are two scalar fields which are sections of different bundles $L_1$ and $L_2$. These two scalar fields 
are actually $\Phi_1$ and $\Phi_2$ appearing in above equations, and the last equation is not coordinate invariant. 
To make the last equation coordinate invariant on a curved Riemann surface, we need to use the fixed holomorphic section of line bundles $L_{1}$ and $L_{2}$, i.e.  $s\in K\bigotimes L_1^{-1}$ and 
$t\in K \bigotimes L_2^{-1}$ and their conjugate $s^*, t^*$ \footnote{In case one of the section is zero, we can twist the gauge fields instead.}, and the last equation becomes
\begin{equation}
F_{z\bar{z}}+[\Phi_{1}, \Phi_1^*]ss^*+[\Phi_2,\Phi_2^*]tt^*=0.
\end{equation}
There are some obvious solutions, i.e. if $\Phi_1=\Phi_2=0$, then the moduli space of above equation is the moduli space of stable holomorphic bundle \cite{atiyah1983yang}. If $\Phi_1$ or $\Phi_2$ equals to zero, one 
get the moduli space of stable twisted Higgs bundle as studied in \cite{nitsure:1991}.

Using the above equations in the local form, one can find the following regular singular boundary condition:
\begin{align}
&\Phi_1={e_1\over z},~~\Phi_2={e_2\over z} \nonumber,~~A_{\bar{z}}={h_1\over \bar{z}}+{h_2\over \bar{z}}, \nonumber\\
&[e_1, e_2]=0,~~~[h_1,h_2]=0, \nonumber\\
&[h_1,e_1]=e_1,~~[h_2,e_2]=e_2,
\end{align}
Here $e_1, e_2$ are nilpotent and $h_1, h_2$ can be taken as semi-simple.
Interestingly such nilpotent pair with the corresponding semi-simple pair has been studied in detail by Ginzburg \cite{ginzburg2000principal}, and we conjecture the above ones are the most general regular singular boundary condition. 
The local Coulomb branch is then described by orbit of nilpotent commuting pair.

Again, the local Higgs branch might be described by moduli space of generalized Nahm's equation. Similarly, such 
BPS equation can be derived by further reducing 6d self-dual equation down to 1d, then we get a generalized Nahm's equation:
\begin{align}
&D_1 \phi_3=[\phi_2,\phi_4],~~D_1 \phi_4=-[\phi_2,\phi_3] \nonumber \\
&D_1 \phi_5=[\phi_2,\phi_6],~~D_1 \phi_6=-[\phi_2,\phi_5] \nonumber \\
&[\phi_3, \phi_5]=[\phi_4,\phi_6]~~~[\phi_3,\phi_6]=[\phi_5,\phi_4]\nonumber \\
&D_1 \phi_2=[\phi_3, \phi_4]+[\phi_5,\phi_6].
\end{align} 
We want to consider singular solutions to above generalized Nahm equation. Let's rewrite the above equations into the real part and complex part by defining
\begin{equation}
 \alpha=A_1-i\phi_2,~~~\beta={1\over2}(\phi_3+i\phi_4),~~~\gamma={1\over2}(\phi_5+i\phi_6 ),
 \end{equation}
 and we have
\begin{align}
& {d\beta \over ds}+[\alpha, \beta]=0,~~{d\gamma \over ds}+[\alpha, \gamma]=0,~~ [\beta, \gamma]=0,~~~~~\text{Complex}  \nonumber\\
& {d (\alpha^*-\alpha)\over ds}+[\alpha, \alpha^*]+[\beta, \beta^*]+[\gamma,\gamma^*]=0,~~~~~~~\text{Real}
\end{align}
Again the imposition of the real equation is equivalent to dividing the complex equation by complex gauge transformation.  Rgular singular solution to 
above complex equation can be easily found:
\begin{equation}
\alpha={h_1^D+h_2^D\over s},~~\beta={e_1^D\over s},~~\gamma={e_2^D\over s},
\end{equation}
here $e_1^D,e_2^D$ are nilpotent elements and $h_1^D, h_2^D$ are semi-simple elements: 
\begin{itemize}
\item $[e_1^D, e_2^D]$=0~~$[h_1^D, h_2^D]=0$, 
\item $[h_1^D, e_1^D]=e_1^D$,~~$[h_2^D, e_2^D]=e_2^D$.
\end{itemize}
Again, the moduli space of generalized Nahm's equation with the above singular boundary condition on $s=0$ and trivial boundary condition at $s=\infty$
describes the local Higgs branch, and the label of commuting nilpotent orbits at $s=0$ should be the transpose (in proper sense) of the ones used in the boundary condition 
of generalized Hitchin's equation.

\begin{center}
\begin{figure}[htbp]
\small
\centering
\includegraphics[width=8cm]{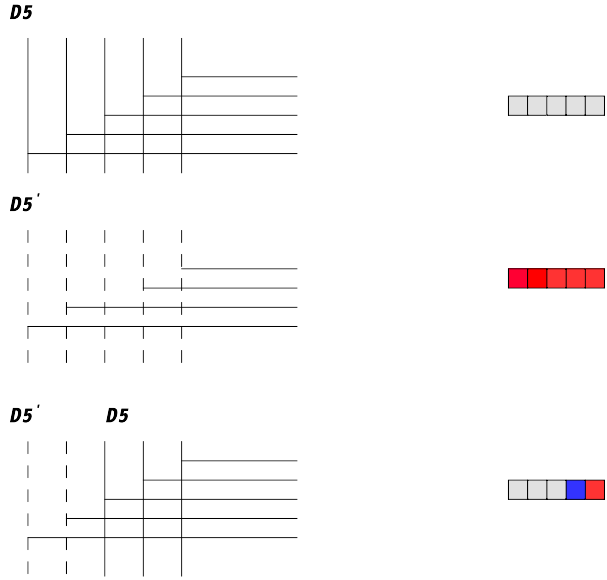}
\caption{$\mathcal{N}=1$ rotated puncture is represented by $D3-D5-D5^{'}$ brane system and labeled by a colored Young Tableaux.}
\label{rotated}
\end{figure}
\end{center}

\begin{center}
\begin{figure}[htbp]
\small
\centering
\includegraphics[width=10cm]{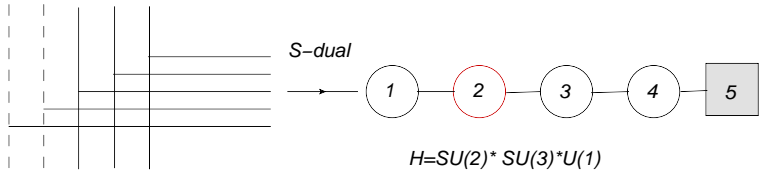}
\caption{The flavor symmetry of a rotated puncture could be read from the Coulomb branch of S dual 3d quiver. Red circle means that we have 3d $\mathcal{N}=2$ gauge group.}
\label{flavor}
\end{figure}
\end{center}

In this paper, we will focus on the puncture which can be 
derived by rotating $\mathcal{N}=2$ punctures: change
 some of $\text{D}5$ branes to $\text{D}5^{'}$ branes by rotating nighty degrees in the brane description of $\mathcal{N}=2$ puncture, and we will have the quarter BPS boundary condition instead of half-BPS boundary
condition for $\mathcal{N}=4$ SYM theory. The study of this set of puncture is already quite rich, and 
we call them rotated puncture.  The puncture is then still labeled by a Young Tableaux with colors as the D3 branes can suspend between different types of D5 branes:
\begin{itemize}
\item A column is colored as black if it is represented by D3 branes suspended between two D5 branes.
\item A column is colored as red if it is represented by D3 branes suspended between two $D5^{'}$ branes
\item A column is colored as blue if it is represented by D3 branes suspended between $D5$ and $D5^{'}$ branes.
\end{itemize}
Some examples of rotated punctures are shown in figure. \ref{rotated}. In particular, we call a puncture $D$ ($\text{D}^{'}$) type if the Young Tableaux is completely black (red).
The flavor symmetry could be read from the 3d quiver  derived from doing S duality of $D3-D5-D5^{'}$ system, i.e. the flavor
symmetry is the symmetry on the Coulomb branch of 3d $\mathcal{N}=2$ quiver. Generically, each quiver node contributes a $U(1)$ factor, and we have 
an enhanced $SU(r)$ symmetry if there is a chain of $(r-1)$ 3d $\mathcal{N}=4$ quiver nodes satisfying the balanced condition: $n_f=2n_c$.
See figure. \ref{flavor} for an example. 

\subsection{$\mathcal{N}=1$ theory from M5 brane}
In summary, 4d $\mathcal{N}=1$ theory derived from compactifying six dimensional $(2,0)$ theory of  G type on a punctured Riemann surface is defined by the following data:
\begin{itemize}
\item A punctured Riemann surface $M_{g,n}$ with various type of punctures labeled by nilpotent commuting pair.
\item A rank two  bundle $L_1\bigoplus L_2$ such that their direct product equal to the canonical bundle.
\end{itemize}
We conjecture that in the IR the theory flows to an interacting SCFT. It is nice to have some junction picture in mind about the local 
breaking and global breaking to $\mathcal{N}=2$ theory. For local breaking, we only change the boundary condition and the NS core is not changed. For the global breaking, 
we use both NS  and $\text{NS}^{'}$ three sphere. Brane picture suggests the following breaking pattern:
since there are extra matter coming from fluctuation of $\text{D}3$ brane suspended between $\text{NS}$ and $\text{D}5^{'}$ brane, the local
breaking introduces extra matter and the cubic superpotential term, which breaks $\mathcal{N}=2$ to $\mathcal{N}=1$, and the rotation also completely changes the theory, i.e.
the Coulomb branch deformations are not there. The global breaking, on the other hand, 
break $\mathcal{N}=2$ vector multiplet to $\mathcal{N}=1$ vector multiplet. 

Let's summarize some simple properties of the 4d theory which can be read from the above defining geometric objects:

a: The flavor symmetry is read from the local punctures, and the mass deformations are encoded locally at the punctures. The complex structure moduli 
of the gauge theory could be identified as the exactly marginal deformation whose detailed form will be discussed in next section.

b: There are $U(1)_{45}\times U(1)_{89}$ R symmetries rotating the fibres of two bundles. In the IR, one linear combination becomes the $R$ symmetry, and 
the other one becomes the flavor symmetry. 

c. The moduli space of generalized Hitchin's equation should describe the Coulomb branch of 4d theory compactified on a circle, and the details
will appear elsewhere.

\begin{center}
\begin{figure}[htbp]
\small
\centering
\includegraphics[width=10cm]{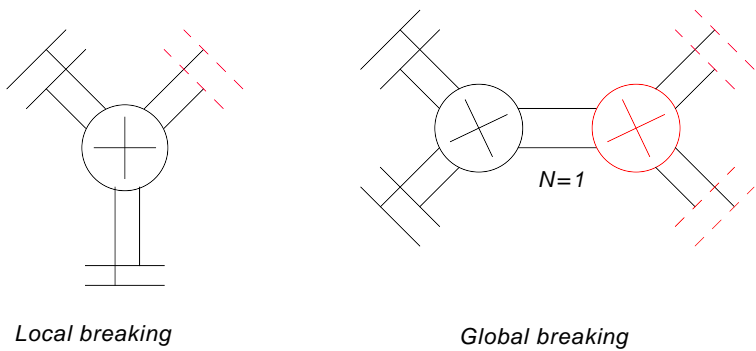}
\caption{Left: Local breaking is achieved by rotating the branes corresponding to punctures. Right: Global breaking is achieved by gluing two different kinds of three sphere. }
\label{one}
\end{figure}
\end{center}

\section{$\mathcal{N}=1$ duality}
Let's consider four dimensional $\mathcal{N}=1$ SCFT derived from 
compactifying 6d $A_{N-1}$ theory on a Riemann surface with $D$ and $D^{'}$ type  punctures.
We are going to interpret the UV gauge theory description as the degeneration limit of the Riemann surface, 
and interpret the duality as taking different degeneration limit of the same Riemann surface.
The natural question would be which coupling constant of the field theory would be identified as the complex 
structure moduli of the Riemann surface.

In  $\mathcal{N}=2$ case, S-duality exchanges the gauge coupling as $\tau\rightarrow {-1\over \tau}$, and 
so if we identify gauge coupling as the complex structure moduli, then S-duality is naturally understood 
as taking different degeneration limits. Notice that the gauge coupling in the conformal case is exactly marginal,
which nicely matches the fact that the complex structure moduli is dimensionless.

By comparing with $\mathcal{N}=2$ S-duality, it is then natural to find the exactly marginal deformations of 
$\mathcal{N}=1$ theory. For $\mathcal{N}=1$ $SU(N)$ SQCD with $2N_f$ flavors, there is indeed an exactly marginal deformation \cite{Leigh:1995ep}: 
\begin{equation}
W=cQ\tilde{Q} Q \tilde{Q}.
\end{equation}
 The gauge group in the Seiberg dual theory is still $SU(N)$ and the superpotential term is 
 \begin{equation}
 W^{'}=Mq\tilde{q}+cMM;
 \end{equation}
after integrating out the massive meson, the superpotentail  becomes:
 \begin{equation}
 W^{'}=-{1\over c} q\tilde{q} q \tilde{q};
 \end{equation}
so under Seiberg duality, $c\rightarrow -{1\over c}$.
Therefore it is suggestive to identify the coupling constant of quartic superpotential term as the complex structure moduli, and 
the Seiberg duality could also be interpreted as different degeneration limits of the Riemann surface.

In general, one can consider quartic superpotential terms preserving certain sub-group of full global symmetries, in the following, we will 
find out what specific superpotential term should be allowed for our M5 brane engineered theories.

\subsection{Insights from type IIA brane construction}
To get some insights about the superpotential term and matter content,
it is useful to review how Seiberg duality works using type IIA brane configurations, then we will find out how Seiberg duality
works in the framework of our M5 brane engineering by lifting the IIA picture to M5 brane description.

\begin{center}
\begin{figure}[htbp]
\small
\centering
\includegraphics[width=10cm]{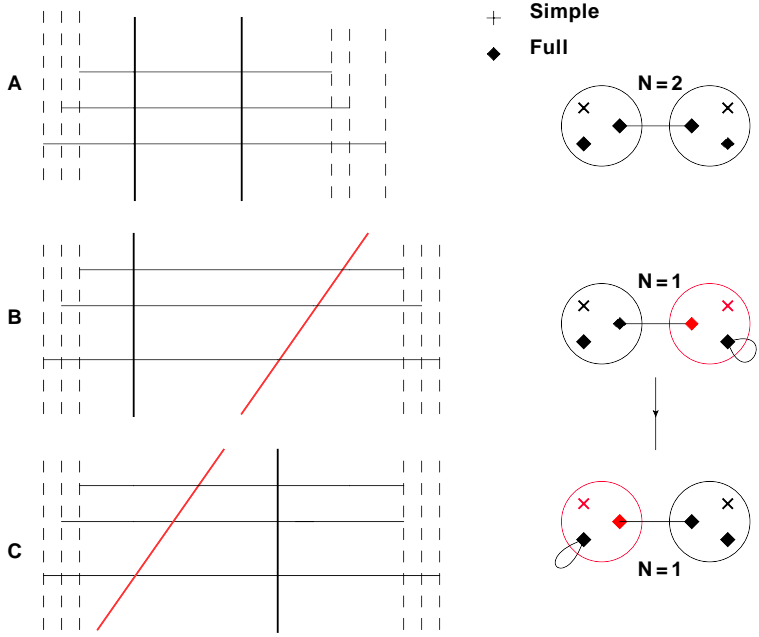}
\caption{Top: Type IIA brane configuration for $\mathcal{N}=2$ SU(N) with $N_f=2N$ and M theory lift. Middle: Type IIA brane configuration for $\mathcal{N}=1$ SU(N) with $N_f=2N$ which is 
achieved by rotating one of NS brane to $\text{NS}^{'}$ brane, and the M5 brane description corresponds to rotating one of simple puncture. Bottom: Seiberg duality is realized as exchanging the position 
of NS and $\text{NS}^{'}$ brane, which corresponds to exchange the position of simple punctures in M5 brane description, we also need to exchange two three spheres. }
\label{lift}
\end{figure}
\end{center}

Let's first start with $\mathcal{N}=2$ theory  engineered using $\text{D}4-\text{D}6-\text{NS}5$ brane system of Type IIA string theory. The summary of
various brane configuration is listed in table. \ref{4d}. The brane set up for  $\mathcal{N}=2$ $\text{SU}(N)$ gauge theory with $N_f=2N$ is shown in figure. \ref{lift}, in which the $2N$ flavors are separated into two equal parts. 
Notice that we use a configuration in which all D6 branes are put on $x_6=\pm \infty$.  

The above brane system is described by $N$ M5 brane wrapping on  a sphere with four punctures: two simple punctures 
describing the intersections of two NS5 branes, and 
two full punctures describing the boundary condition specified by the D6 brane system, see figure. \ref{lift}. The gauge coupling constant is identified with the complex
structure moduli of fourth punctured sphere.

\begin{table}
\begin{center}
  \begin{tabular}{ |l | c |c|c|c|c|c|c|c|c| r| }
    \hline
    ~&$x^0$&$x^1$ & $x^2$& $x^3$&$x^4$&$x^5$&$x^6$&$x^7$&$x^8$&$x^9$ \\ \hline
        D4&$\circ$&$\circ$ & $\circ$& $\circ$&~&~&$\circ$&~&~&~ \\ \hline
    NS5&$\circ$&$\circ$ & $\circ$& $\circ$&$\circ$&$\circ$&~&~&~&~ \\ \hline
    D6&$\circ$&$\circ$ & $\circ$& $\circ$&~&~&~&$\circ$&$\circ$&$\circ$ \\ \hline
    $\text{NS}5^{'}$&$\circ$&$\circ$ & $\circ$& $\circ$&~&~&~&~&$\circ$&$\circ$ \\ \hline
    $\text{D}6^{'}$&$\circ$&$\circ$ & $\circ$& $\circ$&$\circ$&$\circ$&$~$&$\circ$&$~$&$~$ \\ \hline
  \end{tabular}
\end{center}
\caption{Brane configuration for engineering four dimensional $\mathcal{N}=2$ and $\mathcal{N}=1$ theories.}
\label{4d}
\end{table}

Next let's rotate one of $\text{NS}$ brane to $\text{NS}^{'}$ brane, and we get  $\mathcal{N}=1$ $SU(N)$ gauge theory with $N_f=2N$ 
(plus singlets and cubic superpotential, so it is a mixed electric-magnetic theory). 
Now the world volume of $\text{NS}$ brane
is in $x_4, x_5$ direction and $\text{NS}^{'}$ brane is in $x_8, x_9$ direction, and  the isometries $U(1)_{45}$ and $U(1)_{89}$ correspond to two R symmetries of 
 $\mathcal{N}=1$ compactifications.
In M5 brane description, the rotation corresponds to rotating one of  simple puncture from $\text{D}$ type to $\text{D}^{'}$ type, see figure. \ref{lift}, and we also 
represent two matter system as the NS and $\text{NS}'$ three sphere to match the field theory expectation. 
This naive lift of type IIA configuration to  M5 brane configuration matches our general story: the rotated puncture, and $\mathcal{N}=1$ vector multiplet from gluing two different types of
three punctured spheres.

Seiberg duality corresponds to exchanging the positions of NS and $\text{NS}^{'}$ brane \cite{Elitzur:1997fh}, which 
in  M5 brane picture corresponds to exchanging the positions of two simple punctures, and we also need to exchange the $\text{NS}$ and $\text{NS}^{'}$ three sphere at the same time.
So in this sense the Seiberg duality has the same interpretation of $\mathcal{N}=2$ S-duality \cite{Gaiotto:2009we}: it corresponds to exchanging the position of punctures or corresponds to taking different degeneration limits.

We have argued that the complex structure moduli should be identified as the quartic superpotential couplings, but 
the specific form is not known. 
Let's look at the duality in figure. \ref{lift} more closely to determine the exact superpotential term. The electric theory has two sets of quarks $(q^i, \tilde{q}_{\tilde{j}})$ and $(p^k, \tilde{p}_{\tilde{l}})$
with $i=1,\ldots, N$.  There are gauge singlets $M_{k}^{\tilde{l}}$ coming from  $\text{D}4$ brane suspended between D6 and $\text{NS}5^{'}$ brane, and the cubic superpotential between  quark and the meson is 
\begin{equation}
W_1=M_{k}^{\tilde{l}}( p^k\tilde{p}_{\tilde{l}}-{1\over N}tr(p\tilde{p})\delta^k_{\tilde{l}}),
\end{equation}
here we use the fact that the $U(1)$ on the D4 branes is frozen.  Notice that the terms inside the bracket is the momentum map for the flavor symmetry $SU(N)_L$, which we 
denote it as $\mu_L$, and the superpotential can be written as $W=M_L\mu_{L}$
where $M_L$ is the adjoint on $SU(N)_L$.  

In the brane configuration corresponding to Seiberg-dual description, one has the interaction $ W^{'}=M_R\mu_{R}$ from figure. \ref{lift}. 
However, the two theories with the above interactions are not Seiberg-dual to each other, therefore we need to add some more interaction terms to it.  
We propose the following  extra superpotential term  (see \cite{Aharony:1997ju} for related issue)
\begin{equation}
W_2=\lambda tr(\mu_{1}\mu_2)= \lambda \tilde{q}_{~\tilde{j}}^a q^j_{~b}\tilde{p}_{~\tilde{l}}^bp^l_{~a}-{\lambda\over N}tr(p\tilde{p}) tr(q\tilde{q}),
\end{equation}
where $\mu_1$ is the moment map for the $SU(N)$ gauge group action on $(q^i_{~a}, \tilde{q}_{~j}^a)$ , and $\mu_2$ is the moment map for $SU(N)$ gauge group action on $(p^k_{~a}, \tilde{p}_{~l}^a)$.
The full potential for electric theory is then
\begin{equation}
W=W_1+W_2=M_{k}^{\tilde{l}}(p^k\tilde{p}_{\tilde{l}}-{1\over N}tr(p\tilde{p})\delta^k_{\tilde{l}})+\lambda \tilde{q}_{~\tilde{j}}^a q^j_{~b}\tilde{p}_{~\tilde{l}}^bp^l_{~a}-{\lambda\over N}tr(p\tilde{p}) tr(q\tilde{q}).
\end{equation}
After performing Seiberg duality, the fundamentals (anti-fundamentals) become as anti-fundamentals (fundamentals), and new mesons appear, the superpotential changes as
\begin{equation}
W=M_{k}^{\tilde{l}}A^k_{\tilde{l}}+\lambda B^j_{\tilde{l}}C^{l}_{\tilde{j}}+A^k_{\tilde{l}} p^{*}_{k}\tilde{p}^{*\tilde{l}}+D^i_{\tilde{j}} q^{*}_{i}\tilde{q}^{*\tilde{j}}+B^j_{\tilde{l}} q^{*}_{j} \tilde{p}^{*\tilde{l}}+C^{l}_{\tilde{j}} p^{*}_l \tilde{q}^{*\tilde{j}} -{\lambda\over N} trA trD.
\end{equation}
Off diagonal meson and M is massive, and integrate out them out we get the potential
\begin{equation}
W^d=D^{'i}_{\tilde{j}} (q^{*}_i\tilde{q}^{*\tilde{j}}-{1\over N}tr(q^*\tilde{q^*})\delta^k_{l}))-{1\over \lambda} q^{*}_{j} \tilde{p}^{*\tilde{l}}p^{*}_l \tilde{q}^{*\tilde{j}}+{1\over N \lambda} tr(p^{*}\tilde{p*}) tr(q^*\tilde{q^*})
\end{equation}
here $D^{'}$ is the traceless part of $D$, this potential has precisely the same form as the electric theory with the coupling constant reversed: $\lambda\rightarrow -{1\over\lambda}$ !  Therefore, the superpotential of our theory can be written in a simple way:
\begin{equation}
W=ctr\mu_1\mu_2+tr \mu_A M_A 
\end{equation}
where $\mu_1$ and $\mu_2$ are the moment map for two flavor symmetries which are gauged, and
 $M_A$ is the adjoint on flavor symmetry $SU(N)_A$. The exact marginal  coupling constant $c$ might be interpreted as the length of the long tube in the degeneration limit.
 
\begin{center}
\begin{figure}[htbp]
\small
\centering
\includegraphics[width=9cm]{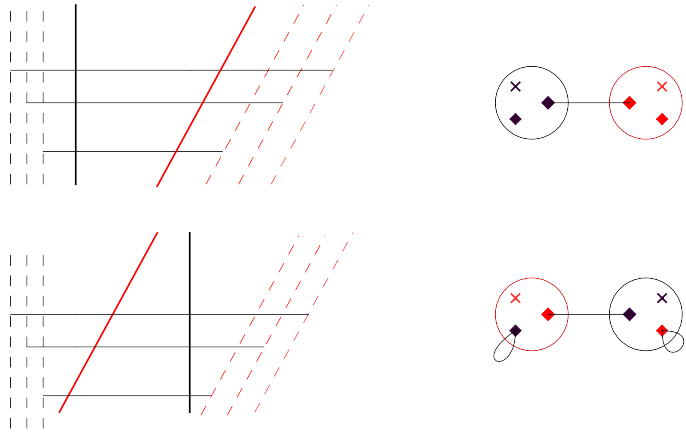}
\caption{A different $\mathcal{N}=1$ theory is described by rotating both NS and D6 branes, and the M theory lift is also shown. Seiberg duality again corresponds to exchange
the position of simple punctures and two types of three spheres.}
\label{lift1}
\end{figure}
\end{center}

One can rotate other branes to get different $\mathcal{N}=1$ theories: for instance, we can rotate D6 and NS branes on far left, which seems to be a good possibility. The brane configuration and its 
M theory representation is shown in figure. \ref{lift1}, and the theory is SQCD without any gauge singlets. Seiberg duality is once again understood as different degeneration limits 
of the same Riemann surface, i.e. by exchanging the positions of two simple punctures and two types of three sphere.
Again, there is  a quartic superpotential term for each of the duality frame to match the brane picture.

\newpage
\subsection{$\mathcal{N}=1$ duality from M5 brane}
\subsubsection{A short review of $\mathcal{N}=2$ duality}
 $S$ duality of 4d $\mathcal{N}=2$ theory corresponds to different degeneration limits of the same punctured Riemann surface (let's assume all the punctures are full for simplicity): 
\begin{itemize}
\item In the complete degeneration limit, there are two new full punctures appearing in degenerating a long tube, which represents a $N=2$ vector multiplet.
\item The Riemann surface is degenerated into several three punctured spheres  representing $T_N$ theory.
 The gauge theory is derived by gauging the  $SU(N)$ flavor symmetries of $T_N$ theory.
\end{itemize}

When the punctures are general, one have the similar story except that newly appearing puncture 
is not the full puncture and three punctured sphere is generically an isolated SCFT.  
The understanding of the S duality is reduced to identify the weakly coupled gauge group and matter system,
 which is answered in \cite{Nanopoulos:2010ga,Chacaltana:2010ks} for $A_{N-1}$ case, and \cite{Chacaltana:2011ze} for $D_N$ theories.

\begin{center}
\begin{figure}[htbp]
\small
\centering
\includegraphics[width=12cm]{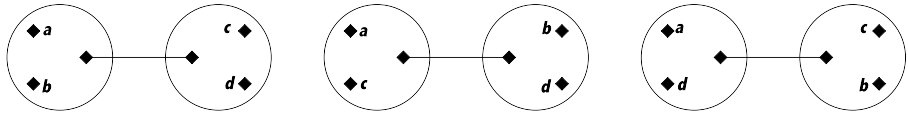}
\caption{S duality for $\mathcal{N}=2$ theory corresponds to different degeneration limit of the same Riemann surface.}
\label{one}
\end{figure}
\end{center}

\subsubsection{$\mathcal{N}=1$ duality: three punctured sphere and gluing}
Motivated by the type IIA brane construction and it's lift to M theory, we would like to conjecture that $\mathcal{N}=1$ duality is realized the same
way as $\mathcal{N}=2$ theory: different duality frames correspond to different degeneration limits of the same punctured Riemann surface whose
defining data is explained in last section.

The first question is that how the global breaking, i.e. different bundle structure is reflected in the degeneration limit? Motivated by the theory without punctures which is 
considered in \cite{Bah:2012dg}, we propose that the global breaking is encoded by having both $\text{NS}$  and $\text{NS}^{'}$ type three spheres whose number is determined by the degree of bundles and 
puncture types. 
Let's assume there are $n_{NS}$ $\text{NS}$ sphere and $n_{NS^{'}}$ $\text{NS}^{'}$ sphere, which might be determined by the following formula:
\begin{equation}
n_{NS}=p+n_{D},~~~n_{NS^{'}}=q+n_{D^{'}},
\label{part}
\end{equation}
where $p=deg(L_1)$ and $q=deg(L_2)$ with $p+q=2g-2$.  

The heuristic interpretation for the above formula is that: 
 the Euler number $2g-2+n$ is separated two parts: $\text{NS}$ ($\text{NS}^{'}$) part is due to the number of $\text{D}$ ($\text{D}^{'}$) type punctures and the line bundle $L_1$ ($L_2$). 
 If we assume each $\text{NS}$ ($\text{NS}^{'}$) three sphere contributes one unit to $\text{NS}$ ($\text{NS}^{'}$) part of Euler number, 
then we got the above formula.
In this paper, we restrict to the case where both $n_{NS}$ and $n_{NS^{'}}$ are non-negative. 

Consider a three sphere with only maximal punctures, then we have the following possibilities

1. $n_{D}=3, n_{D^{'}}=0, n_{NS}=1$, then the line bundle structure is ${\cal O}(-2)\bigoplus{\cal O}(0)$.

2. $n_{D}=3, n_{D^{'}}=0, n_{NS^{'}}=1$, then the line bundle structure is ${\cal O}(-3)\bigoplus{\cal O}(1)$.

3.  $n_{D}=2, n_{D^{'}}=1, n_{NS}=1$, then the line bundle structure is ${\cal O}(-1)\bigoplus{\cal O}(-1)$.

4.  $n_{D}=2, n_{D^{'}}=1, n_{NS^{'}}=1$, then the line bundle structure is ${\cal O}(-2)\bigoplus{\cal O}(0)$.

\begin{center}
\begin{figure}[htbp]
\small
\centering
\includegraphics[width=8cm]{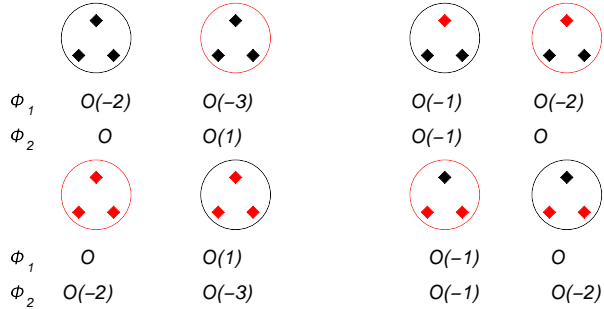}
\caption{The bundle structures of different three punctured sphere with full punctures.}
\label{sphere}
\end{figure}
\end{center}

There are other four type of three spheres derived by exchanging $\text{D}$ type puncture with $\text{D}^{'}$ type puncture, and 
$\text{NS}$ type sphere with $\text{NS}^{'}$ type three sphere, see figure. \ref{sphere}.

We depict the $\text{NS}$ three sphere as black circle and $\text{NS}^{'}$  three sphere as 
red circle. Based on the brane picture studied in last subsection, we can easily find the gauging rules:
\begin{itemize}
\item If two three spheres with same color are glued together, there is a $\mathcal{N}=2$ vector multiplet and usual $\mathcal{N}=2$ coupling:
\begin{equation}
W=\tau tr\Phi( \mu_1-\mu_2).
\end{equation}
\item If two three spheres with opposite color are glued together, there is a $\mathcal{N}=1$ vector multiplet and a superpotential term:
\begin{equation}
W=ctr(\mu_1\mu_2),
\end{equation}
where $\mu_1, \mu_2$ are the moment maps for two gluing punctures. 
\end{itemize}
we also assume that a new D ($\text{D}^{'}$) type puncture appears on the black (red) three sphere in the complete degeneration limit.

\begin{center}
\begin{figure}[htbp]
\small
\centering
\includegraphics[width=8cm]{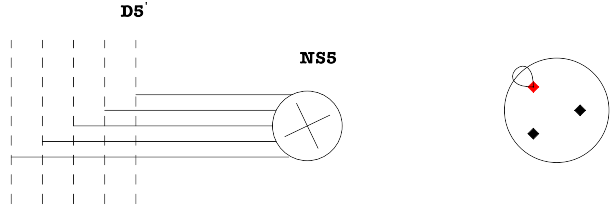}
\caption{There are new gauge singlets coming from the fluctuation of $\text{D}3$ branes suspended between $D5^{'}$ branes and $NS$ brane. }
\label{matter}
\end{figure}
\end{center}

The remaining task is to determine the matter system corresponding to three sphere with rotated puncture. The local effects can be easily seen using the brane construction. Let's assume the three sphere is $\text{NS}$ type, and 
look at a rotated full puncture: a $\text{D}^{'}$ type full puncture. Because of the rotated puncture, one need to use generalized Nahm's equation. The scalar $Y=x_8+ix_9$ now has trivial Nahm pole, and therefore $X=x_4+ix_5$ representing the 
fluctuation for D3 brane suspended between $\text{D}5^{'}$ and NS branes is unconstraint and give a gauge singlet $\text{M}$ which is the adjoint of $SU(N)$ flavor symmetry. We also have the cubic superpotential
 \begin{equation}
W=Tr (\mu_G M),
\end{equation}
here $\mu_G$ is the moment map for flavor symmetry $SU(N)$, see figure. \ref{matter}.

For general rotated punctures specified by a Young Tabeaux and a $SU(2)$ homomorphism $\rho$,  the $D5^{'}$ brane system gives a Nahm pole boundary condition on scalar $Y$, which 
break the flavor symmetry to the commutant $H$, and the moment map for the flavor symmetry is $\mu_H$. From the generalized Nahm's equation, only the $X$ component commuting with $\rho$ is preserved, namely 
\begin{equation}
X\in g^\rho,
\end{equation}
and the superpotential becomes
\begin{equation}
W=Tr( \mu_H X_H).
\end{equation}
For each nilpotent element $Y$, one can associate a 
standard triple $(e,f,h)$ where $e=Y$, and $h$ is a semi-simple element. $h$ provides a grading of the lie algebra $g$ which decomposes as
\begin{equation}
g=\bigoplus g_i
\end{equation}
where $g_i=\{g| [h,g]=ig\}$. Therefore the centralizer of $Y$ is also factorized as 
\begin{equation}
g^Y=\bigoplus g_i^Y
\end{equation}
and the centralizer of the homomorphism $\rho$ is actually given by $g^\rho=g_0^Y$, which gives the reductive group identified with flavor group.

In summary, the local effect of rotating a general puncture is the following: the flavor symmetry is broken to the commutant of $\rho$ and 
there are $g^\rho$ number of gauge singlets, which is equal to the dimension of the flavor group. For example, in the case of full puncture, there are $N^2-1$ gauge singlets,
and in the case of simple puncture, there is only one gauge singlet. Notice that in our treatment, the number of gauge singlets is different from what is found in \cite{Gadde:2013fma}\footnote{They count the number of gauge singlets as $g^Y$, namely the
centralizer of Nilpotent element, here according to our generalized Nahm's equation, the number of gauge singlets is equal to $g^\rho$.}.

\begin{center}
\begin{figure}[htbp]
\small
\centering
\includegraphics[width=8cm]{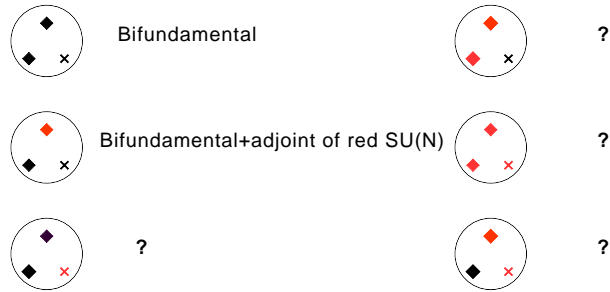}
\caption{Various three punctured sphere theory defined using two full punctures and one simple puncture. In the first and second case, one can express the matter content in terms of free fields using
type IIA brane picture. }
\label{theory}
\end{figure}
\end{center}

However, after rotating the puncture, the three puncture theory is very different from the original there punctured sphere, 
as they are represented by completely different data: different bundles, different puncture types.
In some cases, one can express the new three punctured sphere theory in terms of original $\mathcal{N}=2$ matter content plus the gauge singlet introduced above from local analysis,
but in general, the new theory is not related to the original theory in a simple way, see figure. \ref{theory}, some of them 
is even strongly coupled. In this paper, we simply assume the fact that each $\mathcal{N}=1$ matter system is uniquely specified by a three punctured sphere whose defining data are bundles and puncture types.
It would be really interesting to further study these matter systems.

\subsubsection{Examples}
Using the above rules, we are going to study the dualities of several interesting examples involving punctures. 

\begin{center}
\begin{figure}[htbp]
\small
\centering
\includegraphics[width=10cm]{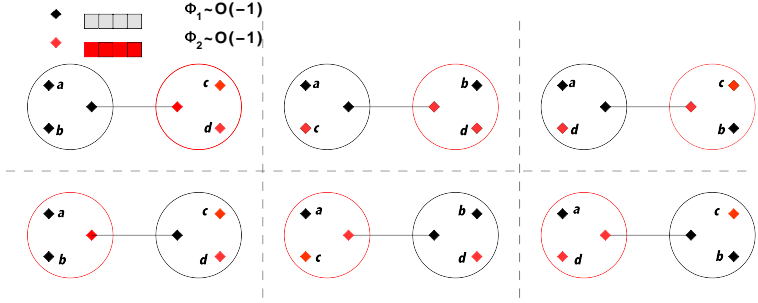}
\caption{Different duality frames of a theory defined by full punctures. }
\label{tn}
\end{figure}
\end{center}

\textbf{Example 1}: Let's consider a sphere with two full D type punctures and two full $\text{D}^{'}$ type punctures, and the line bundle is $L_1=L_2={\cal O}(-1)$, then according to 
formula [\ref{part}], there are one NS sphere and one $\text{NS}^{'}$ sphere.
There are six duality frames as shown in figure. \ref{tn}, and one of them is described by $\mathcal{N}=1$ vector multiplet coupled with two $T_N$ theories, and there is a quartic superpotential.
Notice that the coupling constant is proportional for duality frames in the same column, as 
we only exchange the NS and $\text{NS}^{'}$ type spheres without moving the punctures (without changing the complex structure moduli and therefore the quartic coupling). 

\textbf{Example 2}: $\text{SU}(N)$ SQCD with $N_f=2N$ could be realized by a sphere with four punctures: D type full puncture and simple puncture, plus $\text{D}^{'}$ type 
full and simple puncture. The line bundle structure is also $L_1=L_2={\cal O}(-1)$. In the degeneration limit, there are one $\text{NS}$ and one $\text{NS}^{'}$ sphere. 
There are also six duality frames as discussed in \cite{Gadde:2013fma}, but the matter system in our treatment is quite different, i.e. we do not know if there is a free field representation 
for various three punctured sphere appearing in other duality frames except the standard Seiberg-dual in which the three punctured sphere could be represented by 
a bifundamental fields and gauge singlets transforming as adjoint of one of $SU(N)$ flavor group.
\begin{center}
\begin{figure}[htbp]
\small
\centering
\includegraphics[width=10cm]{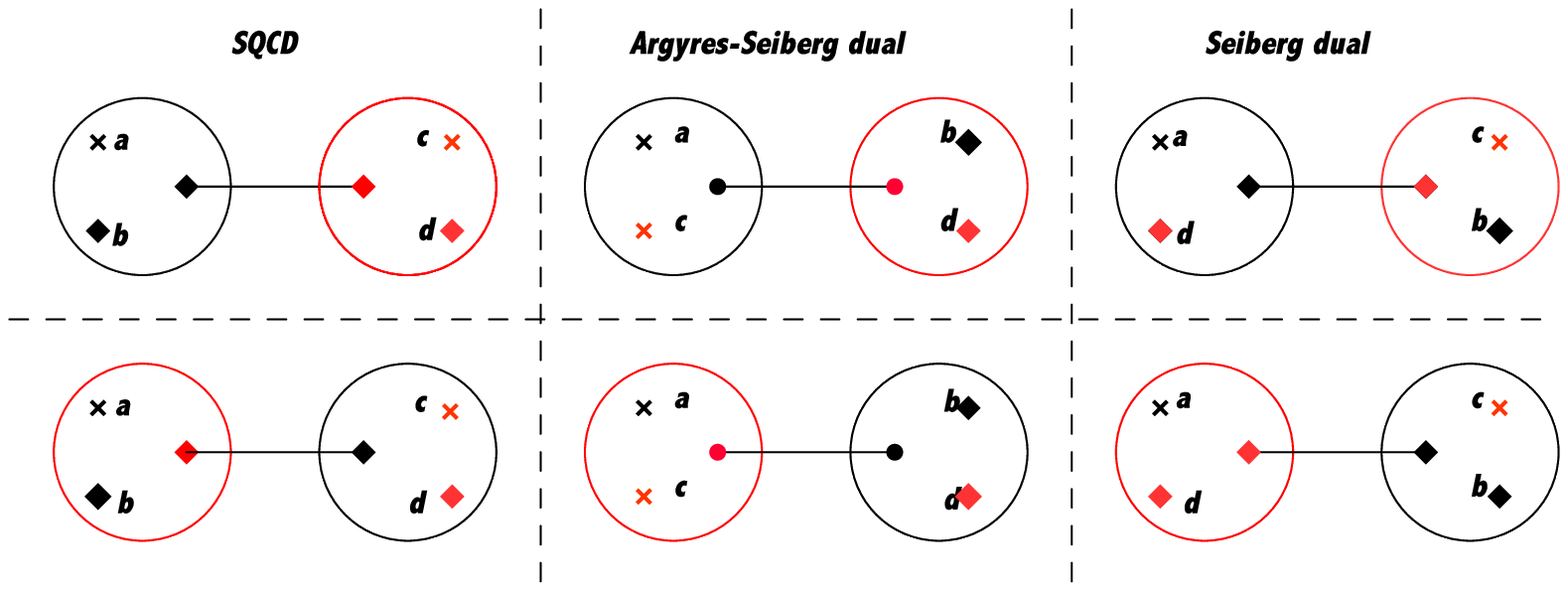}
\caption{Different duality frames of SU(N) SQCD with $N_f=2N$. }
\label{one}
\end{figure}
\end{center}

\textbf{Example 3}: Let's consider a sphere with two D type full punctures, and two D type simple punctures, and one $\text{D}^{'}$ type puncture, and the line bundle structure is ${\cal O}(-2)\bigoplus {\cal O}$, so there are two $\text{NS}$ spheres
and one $\text{NS}^{'}$ sphere. 
In one duality frame, one get a linear quiver with two  $\mathcal{N}=1$ $SU(N)$ gauge groups and quartic superpotential terms. 
In another duality frame, one has $\mathcal{N}=2$ gauge group and various mesons and cubic superpotential term.

\begin{center}
\begin{figure}[htbp]
\small
\centering
\includegraphics[width=10cm]{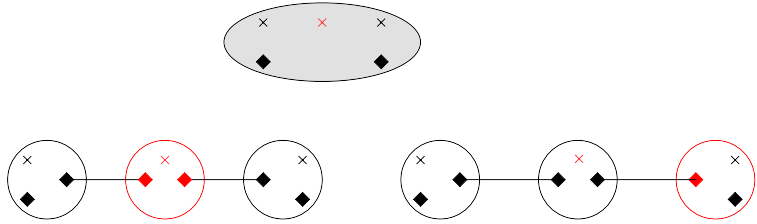}
\caption{A theory is defined by a sphere with five punctures, and we show two duality frames.}
\label{one}
\end{figure}
\end{center}

\begin{center}
\begin{figure}[htbp]
\small
\centering
\includegraphics[width=10cm]{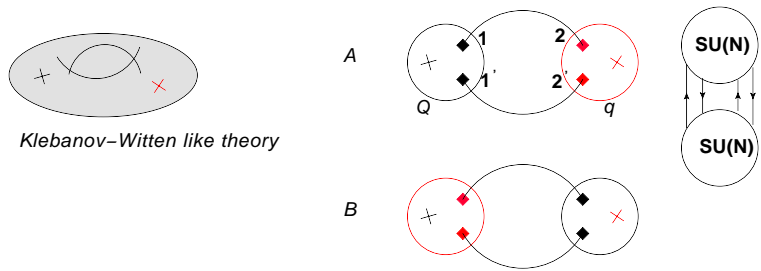}
\caption{The new dual of Klebanov-Witten like theories. }
\label{kw}
\end{figure}
\end{center}

\textbf{Example 4}: Let's consider a torus with one D type simple puncture and one $\text{D}^{'}$ type simple puncture, and the line bundle is $L_1=L_2={\cal O}(-1)$. There are one NS sphere and 
one $\text{NS}^{'}$ sphere in weakly coupled gauge theory description. In one duality frame, we have a Klebanov-Witten \cite{Klebanov:1998hh} like theory with superpotential:
\begin{equation}
W=c_1Tr(\mu_1\mu_2)+c_2Tr(\mu_{1^{'}}\mu_{2^{'}}).
\end{equation}
The conventional Seiberg duality corresponds to exchange both the simple puncture and three spheres. If we only exchange the simple punctures, we find a different duality frame where we
have gauge singlets and cubic superpotential term as shown in figure. \ref{kw}B, which is a new dual of the Klebanov-Witten like theory.

\section{$\mathcal{N}=1$ duality for other theories}

\subsection{Partially rotated puncture}
In last section, we studied  $\mathcal{N}=1$ duality of theories defined using D and $\text{D}^{'}$ type punctures. In this part, we briefly discuss the theory defined using partial rotated puncture.
The story is quite similar: in the degeneration limit,  the number of $\text{NS}$ or $\text{NS}^{'}$ three spheres are given by the formula [\ref{part}], and now $n_{D}$ ($n_{D^{'}}$) means the number
of punctures which have $D5$ ($D5^{'}$) brane as the starting one in their brane construction; and the gluing is also the same as discussed in last section.
The remaining question is to determine the matter content, namely, how many gauge singlets  are there. This question can be similarly answered by looking at the flavor symmetry generated by
this puncture.

For simplicity, let's consider a rotated full puncture which can be represented by a sequence of numbers $(n_1, n_2, \ldots, n_r)$, which means that there are $n_1$ D5 branes followed by 
$n_2$ $D5^{'}$ branes.  Then the flavor symmetry read from $S$ dual brane configuration is 
\begin{equation}
H=S[\sum_i U(n_i)].
\end{equation}
Now suppose the above puncture is put on a $\text{NS}$ three sphere, then there are new fields coming from D3 branes suspended between NS and $\text{D}5^{'}$ branes. So there are adjoints on $\sum_{i\in even} SU(n_i)$ and a 
corresponding cubic superpotential. If the above puncture is put on a $\text{NS}^{'}$ three sphere, then there are adjoints on $\sum_{i\in odd} SU(n_i)$.

\subsection{$D_N$ theories}
It is also straightforward to generalize the study of dualities to $\mathcal{N}=1$ theory engineered using six dimensional $D_N$ theory.  
We already discussed the generalized Hitchin's equation and regular punctures which are valid for any group. 
One also has a brane construction for various $\mathcal{N}=2$ regular punctures  \cite{Gaiotto:2008ak}, and therefore we can also define the 
rotated puncture in similar way.

Here let's consider two theories for illustration: $SO(2n)$ theory with $N_f=4n-4$ 
and $USp(2n-2)$ theory with $N_f=4n$, notice that the above two theories are self-dual (the  dual gauge group has the same rank) under Seiberg duality,
and the quartic superpotential is exactly marginal since the quark has R charge ${1\over2}$.

\begin{center}
\begin{figure}[htbp]
\small
\centering
\includegraphics[width=10cm]{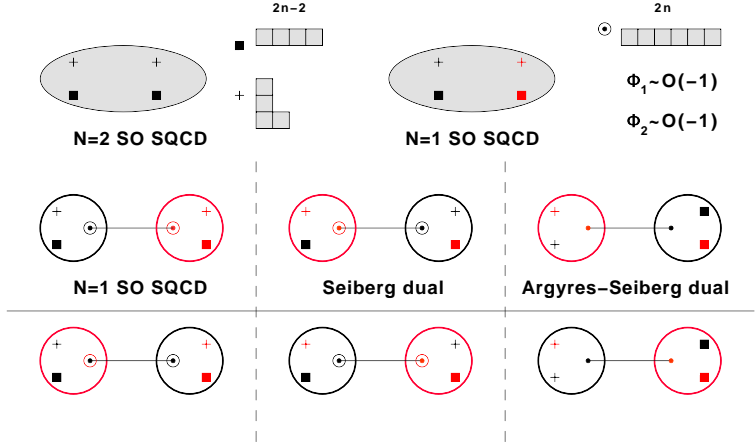}
\caption{The $M5$ brane compactification data for $SO(2n)$ SQCD with $N_f=4n-4$. One find five other duality frames and one of 
them is Seiberg dual. }
\label{so}
\end{figure}
\end{center}

\begin{center}
\begin{figure}[htbp]
\small
\centering
\includegraphics[width=8cm]{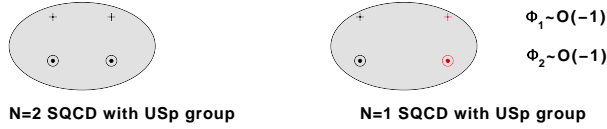}
\caption{The $M5$ brane compactification data for $Usp(2n-2)$ SQCD with $N_f=4n$. One can study dualities by looking at different degeneration limit.}
\label{sp}
\end{figure}
\end{center}

 The M5 brane description for $\mathcal{N}=2$ \cite{Tachikawa:2009rb} and $\mathcal{N}=1$  SO
 SQCD is shown in figure. \ref{so}, here again, we introduce the quartic superpotential term in  $\mathcal{N}=1$ gluing. 
Again, the Seiberg duality is interpreted as taking different degeneration limits of the same Riemann surface, see figure. \ref{so}. 
We also find five other duality frames, and one of them is the standard Seiberg dual. The M5 brane description for USp SQCD is shown in figure. \ref{sp}, and
one can study dualities in exactly same way.

\newpage
\section{Conclusion}
We found two important missing ingredients in studying $\mathcal{N}=1$ theory from M5 brane: a. Generalized Hitchin's equation involving two Higgs 
fields; b. Regular punctures  are classified by orbit of commuting nilpotent pair. We then study $\mathcal{N}=1$ dualities by looking at various degeneration limits of Riemann surface.
There are many open questions:
\begin{itemize}
\item In this paper, we focus on regular punctures which have brane representations.
It is interesting to study more general local puncture i.e. using the algebraic tools developed in \cite{ginzburg2000principal}. The regular puncture 
of $\mathcal{N}=1$ theory is much more fruitful than $\mathcal{N}=2$ theory, for example, the number of regular puncture is infinite, and 
the details will appear in a separate publication \cite{xie:2013uu}.
\item It is interesting to further study the theories defined in this paper, such as the basic three puncture sphere theory,
$a$ maximization, chiral ring, central charges and superconformal index, etc.
The M5 brane construction presented in this paper is expected to be very helpful.
\item It is interesting to study the irregular puncture of $\mathcal{N}=1$ theory, which locally should be classified by commuting first order
differential operator. Using these irregular singularities, one can find a lot of new $\mathcal{N}=1$ Argyres-Douglas type theories \cite{Eguchi:2003wv}, and the details will
appear in \cite{xie:2013uu}. Irregular singularities are also needed for describing confining theories.
\item One can compactify 4d theory on a circle to get 3d $\mathcal{N}=2$ theory, and try to understand 3d Seiberg duality  from 4d duality follow \cite{Aharony:2013dha}, our
M5 brane construction should be very helpful. It is also interesting to learn mirror symmetry of 3d $\mathcal{N}=2$ theory, probably along \cite{Benini:2010uu}.
\item There are some other interesting dynamical question one might be interested to study:
Seiberg-Witten curve on Coulomb branch , phase structure, extended objects such as line operators, surface operators and domain walls,
dynamically generated superpotential \cite{Kazuya:2013}. 
\item The generalized Hitchin's equation plays a key role in our construction, and this equation is not studied before.
Understanding the property of this equation like it's moduli space is perhaps the most important question in trying to learn $\mathcal{N}=1$ gauge dynamics from M5 brane \cite{xie:2013ww}.
\end{itemize}

\begin{flushleft}
\textbf{Acknowledgments}
\end{flushleft}
We thank  Tudor Dimofte, Noppadol Mekareeya, Steve Rayan, Yuji Tachikawa, Masahito Yamazaki  and especially Kazuya Yonekura for help discussions. We would like to thank KIAS for inviting me to
the pre-string conference in which part of the result is presented; We would also like to thank Aspen center for Physics where this paper is finalized.
This research is supported in part by Zurich Financial services membership and by the U.S. Department of Energy, grant DE-SC0009988  (DX).

\appendix

\section{A derivation of local generalized Hitchin's equation}
The local form of the generalized Hitchin's equation can be derived by studying the world volume action of 
$D4$ branes suspended between $D6$ and $D6^{'}$ branes  (so one only consider the fluctuation of the scalar fields in $(x_4, x_5), (x_8, x_9)$ directions), moreover, we assume that
all the fields have only coordinate dependence on $x_1, x_2$, and
the bosonic action  is
\begin{equation}
L=\int d^3 x({1\over4}F_{\mu\nu}F^{\mu\nu}+{1\over2}D_\mu \phi_i D^\mu\phi_i+{1\over4}\sum_{i\neq j}[\phi_i,\phi_j]^2). 
\end{equation}
For our interest, $i=4,5,8,9$. The energy of the static configuration is 
\begin{align}
&E={1\over2}\int d^2 x (F_{12}F_{12}+D_a \phi_i D_a\phi_i+\sum_{i<j}[\phi_i,\phi_j]^2) \nonumber\\
&={1\over2}\int d^2 x[ (F_{12}-[\phi_4,\phi_5]-[\phi_8, \phi_9])^2+(D_1\phi_4-D_2\phi_5))^2+(D_2\phi_4+D_1\phi_5)^2 \nonumber\\
&(D_1\phi_8-D_2\phi_9))^2+(D_2\phi_8+D_1\phi_9)^2+([\phi_4,\phi_8]-[\phi_5,\phi_9])^2+([\phi_4,\phi_9]+[\phi_5,\phi_8])^2+T]
\end{align}
where $T$ is a topological term, so the minimization of energy gives us the following generalized Hitchin's equations
\begin{align}
&D_1 \phi_4=D_2 \phi_5,~~D_1 \phi_5=-D_2\phi_4, \nonumber \\
&D_1 \phi_8=D_2 \phi_9,~~D_1 \phi_9=-D_2\phi_8, \nonumber \\
&[\phi_4, \phi_8]=[\phi_5,\phi_9]~~~[\phi_4,\phi_9]=[\phi_8,\phi_5],\nonumber \\
&F_{12}=[\phi_4, \phi_5]+[\phi_8,\phi_9]. 
\end{align}

\bibliographystyle{utphys} 
 \bibliography{PLforRS}    

\providecommand{\href}[2]{#2}\begingroup\raggedright\begin{thebibliography}{10}

\bibitem{Giveon:1998sr}
A.~Giveon and D.~Kutasov, ``{Brane dynamics and gauge theory},''
  \href{http://dx.doi.org/10.1103/RevModPhys.71.983}{{\em Rev.Mod.Phys.} {\bf
  71} (1999)  983--1084},
\href{http://arxiv.org/abs/hep-th/9802067}{{\tt arXiv:hep-th/9802067
  [hep-th]}}.

\bibitem{Witten:1997sc}
E.~Witten, ``{Solutions of four-dimensional field theories via M-theory},''
  \href{http://dx.doi.org/10.1016/S0550-3213(97)00416-1}{{\em Nucl. Phys.} {\bf
  B500} (1997)  3--42},
\href{http://arxiv.org/abs/hep-th/9703166}{{\tt arXiv:hep-th/9703166}}.

\bibitem{Witten:1997ep}
E.~Witten, ``{Branes and the dynamics of QCD},''
  \href{http://dx.doi.org/10.1016/S0550-3213(97)00648-2}{{\em Nucl.Phys.} {\bf
  B507} (1997)  658--690},
\href{http://arxiv.org/abs/hep-th/9706109}{{\tt arXiv:hep-th/9706109
  [hep-th]}}.

\bibitem{Hori:1997ab}
K.~Hori, H.~Ooguri, and Y.~Oz, ``{Strong coupling dynamics of four-dimensional
  N=1 gauge theories from M theory five-brane},'' {\em Adv.Theor.Math.Phys.}
  {\bf 1} (1998)  1--52,
\href{http://arxiv.org/abs/hep-th/9706082}{{\tt arXiv:hep-th/9706082
  [hep-th]}}.

\bibitem{Argyres:2007cn}
P.~C. Argyres and N.~Seiberg, ``{S-Duality in ${\mathcal{N}}\!=2$
  supersymmetric gauge theories},''
  \href{http://dx.doi.org/10.1088/1126-6708/2007/12/088}{{\em JHEP} {\bf 12}
  (2007)  088},
\href{http://arxiv.org/abs/0711.0054}{{\tt arXiv:0711.0054 [hep-th]}}.

\bibitem{Gaiotto:2009we}
D.~Gaiotto, ``{${\mathcal{N}}\!=2$ dualities},''
\href{http://arxiv.org/abs/0904.2715}{{\tt arXiv:0904.2715 [hep-th]}}.

\bibitem{Maldacena:2000mw}
J.~M. Maldacena and C.~Nunez, ``{Supergravity description of field theories on
  curved manifolds and a no go theorem},''
  \href{http://dx.doi.org/10.1142/S0217751X01003937}{{\em Int.J.Mod.Phys.} {\bf
  A16} (2001)  822--855},
\href{http://arxiv.org/abs/hep-th/0007018}{{\tt arXiv:hep-th/0007018
  [hep-th]}}.

\bibitem{Maldacena:2000yy}
J.~M. Maldacena and C.~Nunez, ``{Towards the large N limit of pure N=1
  superYang-Mills},'' \href{http://dx.doi.org/10.1103/PhysRevLett.86.588}{{\em
  Phys.Rev.Lett.} {\bf 86} (2001)  588--591},
\href{http://arxiv.org/abs/hep-th/0008001}{{\tt arXiv:hep-th/0008001
  [hep-th]}}.

\bibitem{Gauntlett:2003di}
J.~P. Gauntlett, ``{Branes, calibrations and supergravity},''
\href{http://arxiv.org/abs/hep-th/0305074}{{\tt arXiv:hep-th/0305074
  [hep-th]}}.

\bibitem{Tachikawa:2009rb}
Y.~Tachikawa, ``{Six-dimensional $D_N$ theory and four-dimensional SO-USp
  quivers},'' \href{http://dx.doi.org/10.1088/1126-6708/2009/07/067}{{\em JHEP}
  {\bf 07} (2009)  067},
\href{http://arxiv.org/abs/0905.4074}{{\tt arXiv:0905.4074 [hep-th]}}.

\bibitem{Tachikawa:2010vg}
Y.~Tachikawa, ``{N=2 S-duality via Outer-automorphism Twists},''
  \href{http://dx.doi.org/10.1088/1751-8113/44/18/182001}{{\em J.Phys.} {\bf
  A44} (2011)  182001},
\href{http://arxiv.org/abs/1009.0339}{{\tt arXiv:1009.0339 [hep-th]}}.

\bibitem{Chacaltana:2012zy}
O.~Chacaltana, J.~Distler, and Y.~Tachikawa, ``{Nilpotent orbits and
  codimension-two defects of 6d N=(2,0) theories},''
  \href{http://dx.doi.org/10.1142/S0217751X1340006X}{{\em Int.J.Mod.Phys.} {\bf
  A28} (2013)  1340006},
\href{http://arxiv.org/abs/1203.2930}{{\tt arXiv:1203.2930 [hep-th]}}.

\bibitem{Chacaltana:2012ch}
O.~Chacaltana, J.~Distler, and Y.~Tachikawa, ``{Gaiotto Duality for the Twisted
  A(2N-1) Series},''
\href{http://arxiv.org/abs/1212.3952}{{\tt arXiv:1212.3952 [hep-th]}}.

\bibitem{Gaiotto:2009hg}
D.~Gaiotto, G.~W. Moore, and A.~Neitzke, ``{Wall-crossing, hitchin Systems, and
  the WKB approximation},''
\href{http://arxiv.org/abs/0907.3987}{{\tt arXiv:0907.3987 [hep-th]}}.

\bibitem{Nanopoulos:2009uw}
D.~Nanopoulos and D.~Xie, ``{Hitchin equation, singularity, and
  ${\mathcal{N}}\!=2$ superconformal field theories},''
  \href{http://dx.doi.org/10.1007/JHEP03(2010)043}{{\em JHEP} {\bf 03} (2010)
  043},
\href{http://arxiv.org/abs/0911.1990}{{\tt arXiv:0911.1990 [hep-th]}}.

\bibitem{Nanopoulos:2010ga}
D.~Nanopoulos and D.~Xie, ``{$N=2$ Generalized Superconformal Quiver Gauge
  Theory},'' \href{http://dx.doi.org/10.1007/JHEP09(2012)127}{{\em JHEP} {\bf
  1209} (2012)  127},
\href{http://arxiv.org/abs/1006.3486}{{\tt arXiv:1006.3486 [hep-th]}}.

\bibitem{Chacaltana:2010ks}
O.~Chacaltana and J.~Distler, ``{Tinkertoys for Gaiotto duality},''
  \href{http://dx.doi.org/10.1007/JHEP11(2010)099}{{\em JHEP} {\bf 11} (2010)
  099},
\href{http://arxiv.org/abs/1008.5203}{{\tt arXiv:1008.5203 [hep-th]}}.

\bibitem{Xie:2012hs}
D.~Xie, ``{General Argyres-Douglas Theory},''
\href{http://arxiv.org/abs/1204.2270}{{\tt arXiv:1204.2270 [hep-th]}}.

\bibitem{Maruyoshi:2013hja}
K.~Maruyoshi, Y.~Tachikawa, W.~Yan, and K.~Yonekura, ``{N=1 dynamics with TN
  theory},''
\href{http://arxiv.org/abs/1305.5250}{{\tt arXiv:1305.5250 [hep-th]}}.

\bibitem{Benini:2009mz}
F.~Benini, Y.~Tachikawa, and B.~Wecht, ``{Sicilian gauge theories and N=1
  dualities},'' \href{http://dx.doi.org/10.1007/JHEP01(2010)088}{{\em JHEP}
  {\bf 1001} (2010)  088},
\href{http://arxiv.org/abs/0909.1327}{{\tt arXiv:0909.1327 [hep-th]}}.

\bibitem{Bah:2011je}
I.~Bah and B.~Wecht, ``{New N=1 Superconformal Field Theories In Four
  Dimensions},''
\href{http://arxiv.org/abs/1111.3402}{{\tt arXiv:1111.3402 [hep-th]}}.

\bibitem{Bah:2011vv}
I.~Bah, C.~Beem, N.~Bobev, and B.~Wecht, ``{AdS/CFT Dual Pairs from M5-Branes
  on Riemann Surfaces},''
  \href{http://dx.doi.org/10.1103/PhysRevD.85.121901}{{\em Phys.Rev.} {\bf D85}
  (2012)  121901},
\href{http://arxiv.org/abs/1112.5487}{{\tt arXiv:1112.5487 [hep-th]}}.

\bibitem{Bah:2012dg}
I.~Bah, C.~Beem, N.~Bobev, and B.~Wecht, ``{Four-Dimensional SCFTs from
  M5-Branes},'' \href{http://dx.doi.org/10.1007/JHEP06(2012)005}{{\em JHEP}
  {\bf 1206} (2012)  005},
\href{http://arxiv.org/abs/1203.0303}{{\tt arXiv:1203.0303 [hep-th]}}.

\bibitem{Beem:2012yn}
C.~Beem and A.~Gadde, ``{The superconformal index of N=1 class S fixed
  points},''
\href{http://arxiv.org/abs/1212.1467}{{\tt arXiv:1212.1467 [hep-th]}}.

\bibitem{Gadde:2013fma}
A.~Gadde, K.~Maruyoshi, Y.~Tachikawa, and W.~Yan, ``{New N=1 Dualities},''
  \href{http://dx.doi.org/10.1007/JHEP06(2013)056}{{\em JHEP} {\bf 1306} (2013)
   056},
\href{http://arxiv.org/abs/1303.0836}{{\tt arXiv:1303.0836 [hep-th]}}.

\bibitem{ginzburg2000principal}
V.~Ginzburg, ``Principal nilpotent pairs in a semisimple lie algebra 1,'' {\em
  Inventiones mathematicae} {\bf 140} (2000) no.~3, 511--561.

\bibitem{Gaiotto:2008ak}
D.~Gaiotto and E.~Witten, ``{S-Duality of boundary conditions in
  ${\mathcal{N}}\!=4$ Super Yang-Mills theory},''
\href{http://arxiv.org/abs/0807.3720}{{\tt arXiv:0807.3720 [hep-th]}}.

\bibitem{Hashimoto:2013}
A.~Hashimoto, P.~Ouyang, and M.~Yamazaki, ``Work in progress,''.

\bibitem{Elitzur:1997fh}
S.~Elitzur, A.~Giveon, and D.~Kutasov, ``{Branes and N=1 duality in string
  theory},'' \href{http://dx.doi.org/10.1016/S0370-2693(97)00375-4}{{\em
  Phys.Lett.} {\bf B400} (1997)  269--274},
\href{http://arxiv.org/abs/hep-th/9702014}{{\tt arXiv:hep-th/9702014
  [hep-th]}}.

\bibitem{Elitzur:1997hc}
S.~Elitzur, A.~Giveon, D.~Kutasov, E.~Rabinovici, and A.~Schwimmer, ``{Brane
  dynamics and N=1 supersymmetric gauge theory},''
  \href{http://dx.doi.org/10.1016/S0550-3213(97)00446-X}{{\em Nucl.Phys.} {\bf
  B505} (1997)  202--250},
\href{http://arxiv.org/abs/hep-th/9704104}{{\tt arXiv:hep-th/9704104
  [hep-th]}}.

\bibitem{Seiberg:1994pq}
N.~Seiberg, ``{Electric - magnetic duality in supersymmetric nonAbelian gauge
  theories},'' \href{http://dx.doi.org/10.1016/0550-3213(94)00023-8}{{\em
  Nucl.Phys.} {\bf B435} (1995)  129--146},
\href{http://arxiv.org/abs/hep-th/9411149}{{\tt arXiv:hep-th/9411149
  [hep-th]}}.

\bibitem{Leigh:1995ep}
R.~G. Leigh and M.~J. Strassler, ``{Exactly marginal operators and duality in
  four-dimensional N=1 supersymmetric gauge theory},''
  \href{http://dx.doi.org/10.1016/0550-3213(95)00261-P}{{\em Nucl.Phys.} {\bf
  B447} (1995)  95--136},
\href{http://arxiv.org/abs/hep-th/9503121}{{\tt arXiv:hep-th/9503121
  [hep-th]}}.

\bibitem{Bershadsky:1995vm}
M.~Bershadsky, A.~Johansen, V.~Sadov, and C.~Vafa, ``{Topological reduction of
  4-d SYM to 2-d sigma models},''
  \href{http://dx.doi.org/10.1016/0550-3213(95)00242-K}{{\em Nucl.Phys.} {\bf
  B448} (1995)  166--186},
\href{http://arxiv.org/abs/hep-th/9501096}{{\tt arXiv:hep-th/9501096
  [hep-th]}}.

\bibitem{Anderson:2011cz}
M.~T. Anderson, C.~Beem, N.~Bobev, and L.~Rastelli, ``{Holographic
  Uniformization},'' \href{http://dx.doi.org/10.1007/s00220-013-1675-4}{{\em
  Commun.Math.Phys.} {\bf 318} (2013)  429--471},
\href{http://arxiv.org/abs/1109.3724}{{\tt arXiv:1109.3724 [hep-th]}}.

\bibitem{Kapustin:2006pk}
A.~Kapustin and E.~Witten, ``{Electric-Magnetic Duality And The Geometric
  Langlands Program},'' {\em Commun.Num.Theor.Phys.} {\bf 1} (2007)  1--236,
\href{http://arxiv.org/abs/hep-th/0604151}{{\tt arXiv:hep-th/0604151
  [hep-th]}}.

\bibitem{Hitchin:1987ab}
N.~Hitchin, ``{The self-duality equation on a riemann surface},'' {\em
  Proc.London Math.Soc.} {\bf (3)55} (1987)  59--126.

\bibitem{Gukov:2006jk}
S.~Gukov and E.~Witten, ``{Gauge Theory, ramification, and the Geometric
  Langlands Program},''
\href{http://arxiv.org/abs/hep-th/0612073}{{\tt arXiv:hep-th/0612073}}.

\bibitem{Hitchin:1987bc}
N.~Hitchin, ``{Stable bundles and integrable system},'' {\em Duke Math. J.}
  {\bf (1)54} (1987)  91--114.

\bibitem{Gaiotto:2008sa}
D.~Gaiotto and E.~Witten, ``{Supersymmetric boundary conditions in
  ${\mathcal{N}}\!=4$ Super Yang-Mills theory},''
\href{http://arxiv.org/abs/0804.2902}{{\tt arXiv:0804.2902 [hep-th]}}.

\bibitem{Benini:2010uu}
F.~Benini, Y.~Tachikawa, and D.~Xie, ``{Mirrors of 3d Sicilian theories},''
  \href{http://dx.doi.org/10.1007/JHEP09(2010)063}{{\em JHEP} {\bf 1009} (2010)
   063}, \href{http://arxiv.org/abs/1007.0992}{{\tt arXiv:1007.0992 [hep-th]}}.

\bibitem{Corrigan:1982th}
E.~Corrigan, C.~Devchand, D.~Fairlie, and J.~Nuyts, ``{First Order Equations
  for Gauge Fields in Spaces of Dimension Greater Than Four},''
\href{http://dx.doi.org/10.1016/0550-3213(83)90244-4}{{\em Nucl.Phys.} {\bf
  B214} (1983)  452}.

\bibitem{Bak:2002aq}
D.-s. Bak, K.-M. Lee, and J.-H. Park, ``{BPS equations in six-dimensions and
  eight-dimensions},'' \href{http://dx.doi.org/10.1103/PhysRevD.66.025021}{{\em
  Phys.Rev.} {\bf D66} (2002)  025021},
\href{http://arxiv.org/abs/hep-th/0204221}{{\tt arXiv:hep-th/0204221
  [hep-th]}}.

\bibitem{atiyah1983yang}
M.~F. Atiyah and R.~Bott, ``The yang-mills equations over riemann surfaces,''
  {\em Philosophical Transactions of the Royal Society of London. Series A,
  Mathematical and Physical Sciences} (1983)  523--615.

\bibitem{Aharony:1997ju}
O.~Aharony and A.~Hanany, ``{Branes, superpotentials and superconformal fixed
  points},'' \href{http://dx.doi.org/10.1016/S0550-3213(97)00472-0}{{\em
  Nucl.Phys.} {\bf B504} (1997)  239--271},
\href{http://arxiv.org/abs/hep-th/9704170}{{\tt arXiv:hep-th/9704170
  [hep-th]}}.

\bibitem{Chacaltana:2011ze}
O.~Chacaltana and J.~Distler, ``{Tinkertoys for the $D_N$ series},''
  \href{http://dx.doi.org/10.1007/JHEP02(2013)110}{{\em JHEP} {\bf 1302} (2013)
   110},
\href{http://arxiv.org/abs/1106.5410}{{\tt arXiv:1106.5410 [hep-th]}}.

\bibitem{Klebanov:1998hh}
I.~R. Klebanov and E.~Witten, ``{Superconformal field theory on three-branes at
  a Calabi-Yau singularity},''
  \href{http://dx.doi.org/10.1016/S0550-3213(98)00654-3}{{\em Nucl.Phys.} {\bf
  B536} (1998)  199--218},
\href{http://arxiv.org/abs/hep-th/9807080}{{\tt arXiv:hep-th/9807080
  [hep-th]}}.

\bibitem{xie:2013uu}
D.~Xie, ``{M5 brane and four dimensional $N=1$ theory $\text{II}$: general
  punctures, Work in progress},''.

\bibitem{Eguchi:2003wv}
T.~Eguchi and Y.~Sugawara, ``{Branches of N=1 vacua and Argyres-Douglas
  points},'' {\em JHEP} {\bf 0305} (2003)  063,
\href{http://arxiv.org/abs/hep-th/0305050}{{\tt arXiv:hep-th/0305050
  [hep-th]}}.

\bibitem{Aharony:2013dha}
O.~Aharony, S.~S. Razamat, N.~Seiberg, and B.~Willett, ``{3d dualities from 4d
  dualities},''
\href{http://arxiv.org/abs/1305.3924}{{\tt arXiv:1305.3924 [hep-th]}}.

\bibitem{Kazuya:2013}
D.~Xie and K.~Yonekura, ``Work in progress,''.

\bibitem{xie:2013ww}
D.~Xie and Friends, ``Work in progress,''.

\end{thebibliography}\endgroup
\end{document}